\documentclass[useAMS,usenatbib]{mn2e}
\usepackage{epsfig, bm, times, amssymb, graphicx, aas_macros, amsmath}

\usepackage{color}

\newcommand{\del}{{\bm\nabla}}

\newcommand{\curl}{\del\times}
\newcommand{\beq}{\begin{equation}}
\newcommand{\eeq}{\end{equation}}
\newcommand{\beqa}{\begin{eqnarray}}
\newcommand{\eeqa}{\end{eqnarray}}
\newcommand{\bcom}{}

\newcommand{\rosby}{R_{\rm{R}}}
\newcommand{\timt}{\times}

\newcommand{\ddif}{{\rm D}}

\title[A numerical hydrostatic MHD scheme]{A numerical magnetohydrodynamic scheme using the hydrostatic approximation}

\author[J.\ Braithwaite and Y.\ Cavecchi]{Jonathan Braithwaite$^{1}$\thanks{E-mail: jonathan@astro.uni-bonn.de} and Yuri Cavecchi$^{2,3}$\\
$^{1}$Argelander Institut f\"ur Astronomie, Universit\"at Bonn, Auf dem H\"ugel 71, 53121 Bonn, Germany\\
$^{2}$Astronomical Institute ``Anton Pannekoek'', University of
  Amsterdam, Postbus 94249, 1090 GE Amsterdam, The Netherlands\\
$^{3}$Sterrewacht Leiden, University of Leiden,
Niels Bohrweg 2, NL-2333 CA Leiden, The Netherlands}

\pagerange{\pageref{firstpage}--\pageref{lastpage}} \pubyear{}
\begin{document}
\maketitle
\label{firstpage}

\begin{abstract}
In gravitationally stratified fluids, length scales are normally much greater in the horizontal direction than in the vertical one. When modelling these fluids it can be advantageous to use the hydrostatic approximation, which filters out vertically propagating sound waves and thus allows a greater timestep. We briefly review this approximation, which is commonplace in atmospheric physics, and compare it to other approximations used in astrophysics such as Boussinesq and anelastic, finding that it should be the best approximation to use in context such as radiative stellar zones, compact objects, stellar or planetary atmospheres and other contexts. We describe a finite-difference numerical scheme which uses this approximation, which includes magnetic fields.
\end{abstract}

\section{Introduction}
\label{sec:intro}

In magnetohydrodynamical (MHD) simulations, a set of partial differential equations is numerically integrated forwards in time. This is done in stages, with the time increasing in small increments called the ``timestep". When the basic MHD equations are used, the timestep is subject to a range of limits to do with the speed of propagation of information; all numerical explicit schemes become unstable if information is allowed to propagate further than some fraction of a grid spacing in one timestep. For instance, a simple Cartesian hydrodynamical code might have the following timestep:
\begin{equation}
\Delta t = {\rm min} \left[ \frac{C\Delta x}{|u_x| + c_{\rm s}} , \frac{C\Delta y}{|u_y| + c_{\rm s}}, \frac{C\Delta z}{|u_z| + c_{\rm s}} \right]
\label{eq:timestep}\end{equation}
where $\Delta x$, $\Delta y$ and $\Delta z$ are the grid spacings in the three dimensions, ${\bf u}$ is the gas velocity, $c_{\rm s}$ is the sound speed and $C$ is some dimensionless constant whose value will depend on properties of the numerical discretisation scheme, but might be for instance $0.5$. This would ensure that no information can propagate more than $0.5$ grid spacings in any direction during one timestep. Other codes will have additional, similar restrictions from the propagation of Alfv\'en waves and from diffusion; for instance, the sound speed $c_{\rm s}$ in the expression above might be replaced by the fast magnetosonic speed.

By studying the context in which we wish to use simulations, it is often possible to make approximations in order to remove some modes of propagation of information, allowing a larger timestep. Adopting implicit schemes is one efficient way of removing waves. Among the explicit schemes, as well as the basic constant-density incompressible approximation, in which sound and buoyancy waves are both absent, the anelastic \citep{Ogu_Phi:62} and Boussinesq approximations are commonplace \citep[see e.g.][for a review and comparison]{Lilly:96}. Both are widely used in astrophysical hydrodynamics, for instance in studies of convection in planetary and stellar interiors \citep[e.g.][]{Browning:08,Chen_Glatzmaier:05}. They can be used in situations where the thermodynamic variables depart only slightly from a hydrostatically balanced background state, so for instance the density perturbation $\delta\rho/\rho_0\ll 1$. There are some other requirements, such as that the frequency of the motions is much less than the frequency of sound waves and that the vertical to horizontal length scale ratio or the motion is not too large. The two approximations are rather similar, the difference being that the Boussinesq approximation is used where the vertical scale of the motions is much less than the density scale height and where the motion is dominated by buoyancy. It can be shown that the continuity equation, whose standard form is $\partial\rho/\partial t+ {\bm\nabla}\cdot\rho{\bf u}=0$, reduces to the forms ${\bm\nabla}\cdot\rho_0 {\bf u}=0$ and ${\bm\nabla}\cdot{\bf u}=0$ in the anelastic and Boussinesq approximations respectively.
The result of this is that sound waves are filtered out and the timestep is no longer restricted by their propagation. Buoyancy waves are still allowed. 
 
In atmospheric physics, it is common to make the approximation of hydrostatic equilibrium (equation \ref{eq:hydrostatic}), in which we assume perfect vertical force balance. This is applicable in contexts where a constant gravitational field causes strong stratification, where the length scales in the vertical direction are much smaller than in the horizontal, and where the fluid adjusts to vertical force balance on a timescale much shorter than any other timescale of interest -- on the timescale of sound waves propagating in the vertical direction \citep{Richardson:1922}. The consequence is that {\it vertically propagating} sound waves are filtered out, as well as high frequency internal gravity waves, and the $z$-component of the timestep restriction (equation \ref{eq:timestep}) can be removed. This is an obvious advantage in any situation where the vertical length scales present in the system are much smaller than the horizontal length scales and adequate modelling therefore requires that $\Delta z\ll\Delta x$, $\Delta y$. 

The hydrostatic approximation reduces the number of independent variables in the system. For instance, in a system where the gas has two thermodynamic degrees of freedom, in `raw' hydrodynamic equations there are these two thermodynamic variables plus the three components of velocity. In the equivalent hydrostatic system the vertical component of the velocity is no longer independent, but calculated by integration of equation (\ref{eq:hydrostatic}) (see section \ref{sec:equations}); furthermore one of the thermodynamic variables is lost. In the astrophysical context we often want to model conducting fluids with magnetic fields. Note that although the hydrostatic approximation filters out vertically-propagating {\it sound} waves, vertically-propagating magnetic waves are not entirely filtered; a magnetohydrostatic scheme is therefore of use only in the case where the plasma $\beta$ is high, i.e. where the Alfv\'en speed is much less than the sound speed.

There are various ways in which the hydrostatic approximation can be implemented, resulting in different sets of equations and independent variables. The vertical coordinate can be physical height, pressure, entropy or some combination of those \citep[see e.g.][for a review of coordinate systems]{Kasahara:74,Kon_Ara:97}. It turns out, for instance, that a change of the vertical coordinate from height $z$ (as used in other systems) to pressure $P$ simplifies the equations. This can be seen by noting that in hydrostatic equilibrium, the pressure at any point is simply equal to the weight of the column of gas above that point and that each grid box (which has a constant pressure difference $\Delta P$ from top to bottom) will contain constant mass. The continuity equation therefore becomes $\partial u_x/\partial x + \partial u_y/\partial y + \partial \omega/\partial P=0$, where $\omega\equiv \ddif P/\ddif t$ the full Lagrangian derivative, which has the same form as the familiar incompressible equation ${\bm\nabla}\cdot{\bf u}=0$ where vertical velocity $u_z\equiv \ddif z/\ddif t$ has been replaced by $\omega$. Entropy coordinates (also known as isentropic coordinates) are also commonplace, the main advantage being that the vertical `velocity' $\ddif s/\ddif t$ is small, a function only of heating and cooling, which reduces numerical diffusion in the vertical direction.

The best choice of vertical coordinate often depends on the desired upper and lower boundary conditions. In weather forecasting, for instance, it is necessary to have the lower boundary fixed in space. Using pressure coordinates, implementation of this is challenging. It is for this reason that \citet{Kas_Was:67} produced a hydrostatic numerical scheme using height coordinates, but owing to advances in hybrid coordinate systems which allowed also for topographical features -- mountain ranges and so on -- this scheme never became popular. However, when magnetic fields are added, height coordinates $z$ will be simpler than either pressure or entropy coordinates and may regain an advantage in some contexts.

In more astrophysical contexts, such as neutron star, stellar or planetary atmospheres, we may want a lower boundary fixed in space, and to be more precise, fixed at a particular height (unlike in the terrestrial context, mountain ranges and so on need not be included). It is often desirable to have the upper boundary fixed in pressure, if the temperature, and therefore also the pressure scale height, varies by a large factor. For instance, during X-ray bursts on neutron stars the temperature increases by about a factor of ten so that an upper boundary fixed in space would mean insufficient resolution of the relevant layers in cold areas, and extremely low densities and high Alfv\'en speeds.\footnote{It is for this reason that Boussinesq and anelastic schemes are unsuitable here, since they cope with only small variations about a constant reference state.} Entropy coordinates are unsuitable since convection may appear, and in any case the entropy of a co-moving fluid element is expected to change rapidly, removing any advantages of this system. In this context, therefore, the natural choice is the $\sigma$-coordinate system, a pressure-related coordinate first proposed by \citet{Phillips:57}. 
 
In section \ref{sec:equations} we present the basic equations, before describing the finite-difference numerical method in more detail in section \ref{sec:numerical}, presenting simple test cases in section \ref{sec:tests} and summarising in section \ref{sec:conc}.

\section{Basic equations}
\label{sec:equations}

In this section we describe the $\sigma$-coordinate system and how magnetic fields are incorporated.

First of all, we describe the standard MHD equations\footnote{We use c.g.s.~units throughout.} and then go on to the additional equations coming from the hydrostatic approximation. Writing down the horizontal part of the velocity as ${\bf u}$, the horizontal part of the momentum equation is
\begin{equation}
\label{eq:momentum}\rho\frac{\ddif {\bf u}}{\ddif t} = -{\bm\nabla}_{\rm h} P + \frac{1}{4\pi}[{\bm\nabla} \times {\bf B} \times {\bf B}]_{\rm h} + {\bf F}^{\rm visc}_{\rm h}\\
\end{equation}
where the subscript h denotes the horizontal component of a vector, and $\ddif/\ddif t\equiv\partial/\partial t + {\bf u}\cdot{\bm\nabla}_{\rm h} + u_z\partial/\partial z$ is the Lagrangian derivative. ${\bf F}^{\rm visc}$ is the viscous force per unit volume. Writing down the continuity, energy and induction equations and the equation of state (i.e. the perfect gas law), we have
\begin{eqnarray}
\label{eq:continuity}\frac{\partial \rho}{\partial t} \!\!\!&=&\!\!\! -{\bm\nabla}_{\rm h} \!\cdot\! {(\rho \bf u)} - \frac{\partial}{\partial z}(\rho u_z)\\
\label{eq:energy}c_{\rm{P}} \frac{\ddif T}{\ddif t} \!\!\!&=&\!\!\! \frac{1}{\rho} \frac{\ddif P}{\ddif t} + Q,\\
\label{eq:induction}
\frac{\partial{\bf B}}{\partial t} \!\!\!&=&\!\!\! {\bm\nabla} \times ( {\bf u}\times{\bf B} - c {\bf E}^{\rm visc} ) \\
\label{eq:eos}P\!\!\!&=&\!\!\!\rho RT,
\end{eqnarray}
where $u_z$ is the vertical component of the velocity, $Q$ is the heating rate 
 per unit mass
 (including that from heat conduction), $c_{\rm{P}}$ is the specific heat at constant pressure, $R$ is the gas constant (the universal gas constant divided by the mean molecular weight of the gas in question) and other quantities have their usual meanings. 
In this system the vertical coordinate is a quantity $\sigma$ related to the pressure $P$ and the pressure at the top and bottom boundaries $P_{\rm T}$ and $P_{\rm B}$ by
\begin{equation}\label{eq:defofsigma}
\sigma \equiv \frac{P-P_{\rm T}}{P_*} \;\;\;\;\;\;\;\;  {\rm where} \;\;\;\;\;\;\;\; P_\ast\equiv P_{\rm B}-P_{\rm T}.
\end{equation}
One of the important main features of this scheme is that here the upper boundary is fixed at pressure $P_{\rm T}$ (where $\sigma=0$) rather than being fixed at a particular height. The lower boundary at $\sigma=1$ is fixed in space.
Now, the standard set of MHD equations would also contain the $z$-component of the momentum equation (\ref{eq:momentum}) but in the hydrostatic approximation we instead assume that the pressure gradient and gravity are always in perfect balance, i.e. we have the equation of hydrostatic equilibrium
\begin{equation}
\label{eq:hydrostatic}\frac{\partial P}{\partial z} = -g\rho,
\end{equation}
neglecting the vertical component of the Lorentz force -- which is much smaller than the pressure and gravity forces in this high-$\beta$ regime.

In this scheme the fundamental variables which are evolved in time are the horizontal part of the velocity ${\bf u}$, the temperature $T$ and the pressure difference $P_\ast$, plus optional quantities such as heating rate $Q$ which we include here. Note that while $T(x,y,\sigma)$ and other variables are three-dimensional, $P_\ast(x,y)$ is only two-dimensional.

The density, or rather its inverse $\alpha\equiv1/\rho$, is calculated from the equation of state (EOS), which in the ideal gas case is
\begin{equation}\label{eq:eosalpha}
\alpha = \frac{RT}{\sigma P_\ast+P_{\rm T}}.
\end{equation}
This can easily be modified to more complex EOS, for example around the transition between ideal gas and degenerate (electron) gas. 
As before, the vertical component of the momentum equation is replaced by the equation of hydrostatic equilibrium (\ref{eq:hydrostatic}) which takes the form
\beq\label{eq:sigma-hs}
g \,\frac{\partial z}{\partial\sigma} = -\alpha P_\ast, 
\eeq
which we integrate from the lower boundary upwards to give height $z$ and potential $\phi$:
\begin{equation}\label{eq:z}
\phi = gz = P_\ast \int_\sigma^1 \!\alpha\, {\rm d}\sigma^\prime.
\end{equation}
In some sense the quantity $P_\ast$ can be considered a `pseudodensity' as it takes the role of density in the equations -- the continuity equation is:
\begin{equation}\label{eq:sigma-cont}
\frac{\partial P_\ast}{\partial t} + {\bm\nabla}_\sigma \cdot (P_\ast {\bf u}) + P_\ast \frac{\partial \dot{\sigma}}{\partial\sigma} = 0,
\end{equation}
where ${\bm\nabla}_\sigma$ represents the gradient at constant $\sigma$ (unlike ${\bm\nabla}_{\rm h}$ which is at constant height). This equation has an obvious similarity to the familiar form $\partial\rho/\partial t + {\bm\nabla}\cdot (\rho{\bf u}) = 0$. The equivalent of the vertical component of the velocity is $\dot{\sigma}\equiv \ddif\sigma/\ddif t$. Note that $P_\ast$ is not a function of $\sigma$ and so comes outside of the derivative in the third term above. Given the boundary conditions that $\dot{\sigma}=0$ at both upper and lower boundaries, we can integrate equation (\ref{eq:sigma-cont}) from $\sigma=0$ downwards to give
\begin{equation}\label{eq:234}
\sigma\frac{\partial P_\ast}{\partial t} + I + P_\ast \dot{\sigma} = 0 \;\;\;\;\; {\rm where} \;\;\;\;\; I\equiv \int^\sigma_0\!\!{\bm\nabla}_\sigma \!\cdot\!(P_\ast{\bf u}) \,{\rm d}\sigma^\prime,
\end{equation}
so that integrating to the lower boundary $\sigma=1$ gives a predictive equation for $P_\ast$:
\begin{equation}\label{eq:dPstardt}
\frac{\partial P_\ast}{\partial t} =- I_{\sigma=1}.
\end{equation}
Substituting this back into equation (\ref{eq:234}) allows calculation of the vertical velocity $\dot{\sigma}$:
\begin{equation}\label{eq:sigmadot}
P_\ast \dot{\sigma} = \sigma I_{\sigma=1} - I.
\end{equation}

Taking the Lagrangian derivative of the definition of $\sigma$ (equation \ref{eq:defofsigma}) and multiplying by $P_\ast$ gives
\begin{equation}\label{eq:sigmadot2}
P_\ast \dot{\sigma} = \frac{\ddif P}{\ddif t} - \sigma\frac{\ddif P_\ast}{\ddif t},
\end{equation}
where the Lagrangian derivative is defined in the usual way
\begin{equation}
\frac{\ddif}{\ddif t} \equiv \frac{\partial}{\partial t} + {\bf u}\cdot{\bm\nabla}_\sigma + \dot{\sigma}\frac{\partial}{\partial\sigma}
\end{equation}
which we can also use to express the time derivative of $P_\ast$ as
\begin{equation}
\frac{\ddif P_\ast}{\ddif t} = \frac{\partial P_\ast}{\partial t} + {\bf u}_{\sigma=1}\cdot{\bm\nabla}P_\ast.
\end{equation}
Substituting this, equations (\ref{eq:dPstardt}) and (\ref{eq:sigmadot}) into equation (\ref{eq:sigmadot2}) gives us an expression for $\ddif P/\ddif t$:
\begin{equation}
\frac{\ddif P}{\ddif t} = \sigma {\bf u}_{\sigma=1}\cdot{\bm\nabla}P_\ast - I
\end{equation}
which we need to evaluate the time derivative of temperature from the thermodynamic equation
\begin{equation}
c_{\rm P} \frac{\ddif T}{\ddif t} = \alpha \frac{\ddif P}{\ddif t} + Q,
\end{equation}
where $Q$, the rate of heating per unit mass, includes all heating, cooling and conductive terms.

What remains now is the (horizontal part of the) momentum equation:
\begin{equation}
\label{equ:fullmomentum}
\frac{\ddif{\bf u}}{\ddif t} = -{\bm\nabla}_\sigma \phi - \sigma\alpha{\bm\nabla}P_\ast - 2 {\bf\Omega}\times{\bf u} + {\bf F}_{\rm visc} +  {\bf F}_{\rm Lor}
\end{equation}
where $ {\bf F}_{\rm Lor}$ is the horizontal part of the Lorentz force
(see below). It is not immediately obvious how best to go about adding
magnetic fields to this scheme, since calculating real-space gradients
in the vertical direction necessitates first calculating the gradient
$\partial/\partial\sigma$ and then multiplying by
$\partial\sigma/\partial z=-g/\alpha P_\ast$, and also because the
grid points themselves are moving in the vertical direction.

The best way is to start by making a switch of the independent variable from ${\bf B}=(B_x,B_y,B_z)$ to
\beq
\label{equ:magfieldstar}
{\bf B^\ast}\equiv\left(B_x\frac{\partial z}{\partial\sigma},B_y\frac{\partial z}{\partial\sigma},B_z\right),
\eeq
the equations can be significantly simplified, mainly since the time derivative $(\partial z/\partial t)_\sigma$ is not required; what we do require is just the derivative $\partial z/\partial\sigma$, which we have already from equation (\ref{eq:sigma-hs}). Further defining  ${\bm\nabla}^\ast=(\partial/\partial x,\partial/\partial y,\partial/\partial\sigma)$ and ${\bf u}^\ast=(u_x,u_y,{\dot \sigma})$, we have
\begin{equation}
\left(\frac{\partial{\bf B}^\ast}{\partial t}\right)_{x, y, \sigma} = {\bm\nabla}^\ast\times({\bf u}^\ast\times{\bf B}^\ast \;-\; {\bf E}^\ast_{\rm visc}).
\end{equation}
This system ensures conservation of flux. Details of the viscous part of the electric field are given in section \ref{sec:magdiff}.

The Lorentz force is calculated by first dividing the $x$ and $y$ components of ${\bf B}^\ast$ by $\partial z/\partial \sigma$ to find the actual magnetic field ${\bf B}$, then finding the current ${\bf J}$ from $\curl{\bf B}$ in the usual way (where the vertical derivatives are of the form $(\partial \sigma/\partial z)\partial/\partial\sigma$, and simply taking the cross-product of the current with the magnetic field, i.e.
\beq
 {\bf F}_{\rm Lor} = ((\frac{\partial}{\partial x},\frac{\partial}{\partial
 y},\frac{\partial\sigma}{\partial z}\frac{\partial}{\partial \sigma})\times
 \bm{B})\times \bm{B}.
\eeq 

\section{Numerical implementation}\label{sec:numerical}

We now describe how the above equations are integrated numerically, describing the grid, timestepping and the diffusion scheme.

\subsection{Numerical grid}\label{sec:grid}

\begin{figure}
\hspace{0.3cm}\includegraphics[width=0.93\hsize,angle=0]{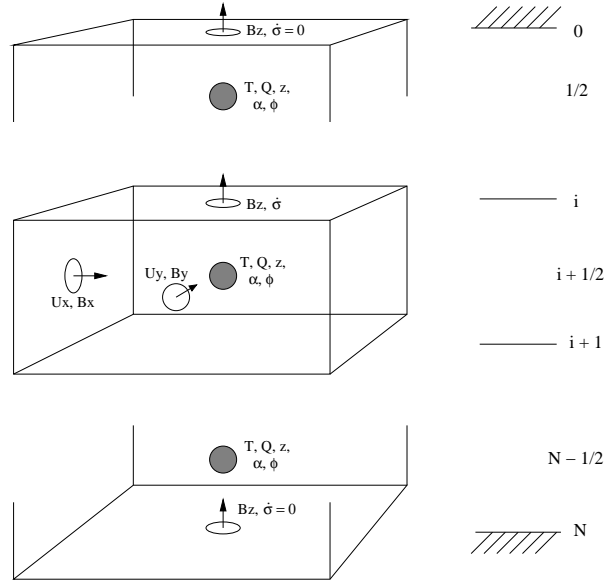}
\caption{The grid, showing the positions within each grid box at which
  the fundamental quantities are defined: $T$, centres of grid boxes,
  dark circle; $U_x$, $\sigma$ and $y$ centred, $x$ face centred, open
  circles; $U_y$, $\sigma$ and $x$ centred, $y$ face centred, open
  circles. $P_\ast(x,y)$ is defined at the same $x$ and $y$ locations
  as $T$; any additional thermodynamic variables such as the heating
  rate $Q$, are defined in the same locations as $T$. $B_x$ is defined
  at the same positions as $U_x$; $B_y$ at the same as $U_y$. $B_z$
  and $\dot\sigma$ are $x$ and $y$ centred, but face centred in the
  $\sigma$ direction. The arrows represent positive directions for all
  the respective vector Kittie's, but $\dot\sigma$: it has the
  opposite sign. See text for definitions. Above and below the central
  box, the upper and lower boundaries are also shown.}
\label{fig:lorenz-grid}
\end{figure}

\begin{figure}
\hspace{0.9cm}\includegraphics[width=0.8\hsize,angle=0]{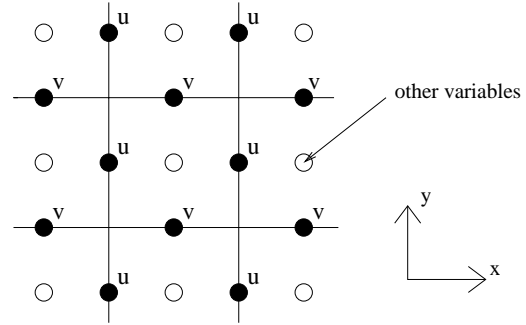}
\caption{The so-called C-grid. Projection on the horizontal
  plane. Velocities are face-centred (dark circles); the open circles
  in the centre of each grid box represent all of the other
  non-magnetic variables. Note that this diagram gives no information
  regarding the vertical positions.}
\label{fig:c-grid}
\end{figure}

The grid is staggered, which improves the conservation properties of the code and is worth the modest extra computational expense; different variables are defined in different positions in the grid boxes. The horizontal components of the velocity are face-centred, defined half of one grid spacing from the centre of each grid box (see Figs. \ref{fig:lorenz-grid} and \ref{fig:c-grid}), whilst all other variables are defined either in the centre of the grid box or vertically above/below it. Temperature and any other thermodynamic variables such as $Q$, and potential $\phi$ are defined at the centre of each grid box\footnote{The staggering in the vertical is known as the {\it Lorenz grid} \citep{Lorenz:60}. The alternative is the {\it Charney-Phillips grid} \citep{Cha_Phi:53} where these quantities are face-centred, displaced half a grid spacing in the vertical from grid-box centre.}. $P_\ast$ is a function of $x$ and $y$ only and is defined in the centre of the grid box, not displaced in the $x$ or $y$ directions. $B_x$, $B_y$ and $B_z$ are face-centred, defined half of a grid spacing from the box centre in the $x$, $y$ and $z$ directions respectively.

At various times while evaluating the time derivatives in the partial differential equations given above,
 it is necessary to find the spatial derivatives of various quantities, to evaluate a quantity at a position other than that where it is defined (e.g. half a grid spacing displaced in some direction) and to integrate a quantity in the vertical direction. The derivatives, interpolations and integrals are evaluated to fifth, sixth and fifth
 order respectively, meaning that the values of the given quantity at six grid points are used to calculate the required quantity/derivative/integral at each grid point \citep[for details see][]{Lele:92}.

For instance, if a quantity $f$ is defined displaced half a grid-spacing in the $x$-direction from the centre of the grid box and its value is required at the grid-box centre the value is calculated thus:
\beqa
f_i =  & &\!\!\!\!\!\!\!\!\frac{75}{128}(f_{i+\frac{1}{2}}+f_{i-\frac{1}{2}})\nonumber\\ & &- \frac{25}{256}(f_{i+\frac{3}{2}}+f_{i-\frac{3}{2}}) + \frac{3}{256}(f_{i+\frac{5}{2}}+f_{i-\frac{5}{2}}),
\eeqa
and if the spatial derivative with respect to $x$ is required at the same location, it is calculated thus:
\beqa
\Delta x\,f'_i = & &\!\!\!\!\!\!\!\!\frac{225}{192}(f_{i+\frac{1}{2}}-f_{i-\frac{1}{2}})\nonumber\\ & & - \frac{25}{384}(f_{i+\frac{3}{2}}-f_{i-\frac{3}{2}}) + \frac{3}{640}(f_{i+\frac{5}{2}}-f_{i-\frac{5}{2}}).
\eeqa
Note that this method of calculating interpolations and derivatives is also used in the `stagger code' \citep{Nor_Gal:1995,Gud_Nor:2005}.

In addition to this, however, we need to integrate various quantities in the vertical direction -- for instance to integrate a body-centred quantity $f$ (ind ices 1/2, 3/2, etc.) from the upper boundary (where coordinate $\sigma=0$) downwards to an arbitrary position $\sigma$, and to return the result at face-centred locations (indices 0, 1, 2 etc.) we first perform a first-order integration
\beq
I^{1{\rm st}}_i = I^{1{\rm st}}_{i-1} + \Delta \sigma\,f_{i-\frac{1}{2}}
\eeq
and then increase the order with the following operation to add parts onto either end of the integration:
\beqa
I_i = I^{1{\rm st}}_i + b(f_{-\frac{1}{2}}-f_\frac{1}{2}) + c(f_{-\frac{3}{2}}-f_\frac{3}{2}) + d(f_{-\frac{5}{2}}-f_\frac{5}{2}) \nonumber\\
                   + b(f_{i+\frac{1}{2}}-f_{i-\frac{1}{2}}) + c(f_{i+\frac{3}{2}}-f_{i-\frac{3}{2}}) + d(f_{i+\frac{5}{2}}-f_{i-\frac{5}{2}}) 
\eeqa
where $b=0.0543134\Delta \sigma$, $c=-0.00484768\Delta \sigma$ and $d=0.000379257\Delta \sigma$. 

In addition, a body-centred integrand must sometimes be integrated from the upper boundary ($\sigma=0$) to body-centred positions. As before, a first-order result is obtained first:
\beq
I^{1{\rm st}}_{i+\frac{1}{2}} = I^{1{\rm st}}_{i-\frac{1}{2}} + \frac{\Delta \sigma}{2}\,(f_{i-\frac{1}{2}}+f_{i+\frac{1}{2}})
\eeq
and a higher order result for just the point half a grid-spacing from the upper boundary:
\beqa
I_{\frac{1}{2}} \!\!\!\!&=&\!\! a_1 f_{\frac{1}{2}}\;\;+\;b_1 f_{\frac{3}{2}}\;\,+\;c_1 f_{\frac{5}{2}}\nonumber\\& &\!\!\!\!\!\!+a_2 f_{-\frac{1}{2}}+b_2 f_{-\frac{3}{2}}+c_2 f_{-\frac{5}{2}}.
\eeqa
where the coefficients have the values 
\begin{align*}
a_1=&&\phantom{-}0.&41410590\Delta\sigma   &  a_2=&&\phantom{-}0.&14283854\Delta\sigma\\
b_1=&&-0.&036306424\Delta\sigma            &  b_2=&&-0.&028276910\Delta\sigma\\
c_1=&&\phantom{-}0.&0041015625\Delta\sigma &  c_2=&&\phantom{-}0.&0035373264\Delta\sigma
\end{align*}
These are used to produce the final result:
\beqa
I_{i+\frac{1}{2}} \!\!\!\!&=&\!\!\!\! I^{1{\rm st}}_{i+\frac{1}{2}} + I_{\frac{1}{2}} - a f_{\frac{1}{2}} \nonumber\\
& &\!\!\!\!\!- b (f_{\frac{3}{2}}\;-\,f_{-\frac{1}{2}})\;-\; c (f_{\frac{5}{2}}\;\,-\,f_{-\frac{3}{2}}) \nonumber\\
& &\!\!\!\!\!+ b (f_{i+\frac{3}{2}}-f_{i-\frac{1}{2}})+ c (f_{i+\frac{5}{2}}-f_{i-\frac{3}{2}})
\eeqa
with $a=(1/2)\Delta\sigma$, $b=-(41/720)\Delta\sigma$ and $c=(11/1440)\Delta\sigma$.
In other situations, it is convenient to perform this integration in the other direction, in which case the equivalent can be done, running the loops in the opposite order.

It can be seen above that for interpolations, derivatives and integrations, the values of quantities are required beyond the boundaries of the computational domain; this is described below in section \ref{sec:boundaries}.

\subsection{Boundaries}\label{sec:boundaries}

In the horizontal directions, the simplest boundaries to implement are obviously periodic boundaries, but there is the possibility of switching one of the two directions to some other boundary condition. Another possibility is to have `mirror' conditions in one direction, in order to avoid modelling the same thing twice in symmetric configurations. These conditions are symmetric in $T$ and other scalars, and antisymmetric in the perpendicular component of the velocity.

In the vertical direction, periodic boundaries are impossible, so symmetric conditions are used in $T$ (and other thermodynamic variables) and the parallel components of velocity. These represent all of the independent variables, as the vertical component of velocity is a derived quantity. The zero condition (and antisymmetry) for the vertical velocity ensures that no mass can flow across the vertical boundaries.

As for the magnetic field, mathematically our boundaries are no
different from the ``pseudo-vacuum'' boundaries used by many other
researchers, the difference is just we evolve ${\bf B^\ast}$ using
${\bf \nabla^\ast}$ and ${\bf u^\ast}$ which have different meanings
(see equation \ref{equ:magfieldstar}), but the time derivative is
still calculated as the curl of a vector. Our quantity ${\bf
  \nabla^\ast}.{\bf B^\ast}$ is therefore conserved just as well as
${\bf\nabla}.{\bf B}$ in other codes, as are the fluxes.

\subsection{Timestepping}\label{sec:timestepping}

The timestepping uses a third-order Low-Storage Runge-Kutta scheme \citep{Williamson:80}, which in practice means that during each timestep the time derivatives on the left-hand sides of the partial differential equations are evaluated three times, with three different values of the quantities on the right-hand sides.
 If the quantities $f$, $g$, etc. are to be evolved in time from time $t=t_0$, at which the quantities have values $f_0$, $g_0$, etc., each time step consists of the following steps (just including variable $f$ for brevity):
\begin{itemize}
\item Calculates time derivatives $f'_0$ from $f_0$
\item Finds timestep $\Delta t$ according to various Courant conditions (see below)
\item Evaluates new values $f_1$
\begin{equation*}f_1=f_0+\Delta t \,b_1 f'_0 \end{equation*}
\item Evaluates new time derivatives $f'_1$ from $f_1$ and averages with previous one
\begin{equation*}f'_{1.5}=a_2 f'_0 + f'_1 \end{equation*}
\item Updates new values for second step
\begin{equation*}f_2=f_1+\Delta t \,b_2 f'_{1.5} \end{equation*}
\item Calculates time derivatives $f'_2$ from $f_2$ and averages with previous ones
\begin{equation*}f'_{2.5}=a_3 f'_{1.5} + f'_2 \end{equation*}
\item Calculates final time step values $f_3$
\begin{equation*}f_3=f_2+\Delta t \,b_3 f'_{2.5}\end{equation*}
\end{itemize}
and so $f_3$ is the value at the end of the timestep. The coefficients are $a_2=-0.641874$, $a_3=-1.31021$, $b_1=0.46173$, $b_2=0.924087$ and $b_3=0.390614$.
The advantage of this type of scheme is that the results from the previous evaluations need not be stored in memory, therefore making the code less demanding in terms of memory usage and faster on systems with limited amount of ram.

The time step is limited by the various Courant conditions given by
the different quantities that are evolved. First define \beqa
\Delta s &\equiv& \min(\Delta x,\Delta y,\Delta\sigma/\sqrt{\nu_\sigma})\\
A&\equiv& (6.2\times3/2)\frac{\Delta t}{\Delta s^2} \eeqa The
coefficients in $A$ come from the following: the $6.2$ from the
maximum value of the second derivative with this sixth-order scheme,
the $3$ from the worst-case 3-D chequered scenario, the $1/2$ to
normalise (since there is a factor $2$ in the diffusive term); see for
instance \citet{art-2009-mar-maclo} and references therein.

Then, we have the following Courant parameters 
\beqa
C_{\rm u} &=& \max(c_{\rm s}+|{\bf u}|)\frac{\Delta t}{\Delta s}\\
C_{\rm p} &=& \max\left(\left|\frac{1}{T}\frac{\partial T}{\partial t}\right|\right) \Delta t\\
C_\nu &=& A\max[3\max(\nu),\max(\nu_{\rm s})] 
\eeqa 
$C_{\rm u}$ is the
limiting factor from velocities (both physical and sound velocity
$c_{\rm s}$), $C_{\rm p}$ the one from temperature changes, while
$C_\nu$ is the one from kinetic diffusion (see
sec. \ref{sec:diffusion}).  Finally, the timestep is fixed according
to \beq \Delta t = \frac{C_{\Delta t}\Delta t}{\max(C_{\rm
    u},C_\nu,C_{\rm p})} \eeq where $C_{\Delta t}$ is some numerical
factor, which we generally set to $0.3$.

\subsection{Diffusion}\label{sec:diffusion}

In this section we describe how the scheme handles kinetic, thermal and magnetic diffusivities.

The code includes both pure physical diffusion as well as a `hyperdiffusive' scheme, designed to damp structure close to the Nyquist spatial frequency while preserving well-resolved structure on larger length scales. Often it is possible, by assessment of their relative magnitudes (also treating the horizontal and vertical directions separately) to switch off one or the other.
 The kinetic, thermal and magnetic diffusivities $\nu$, $\kappa$ and $\eta$ all have dimensions of length squared over time.

It can be shown that with the physical `textbook' diffusion equations, the computational demands sometimes become prohibitively expensive. For instance, using a kinetic diffusivity $\nu$ high enough to handle shocks would result in unacceptable damping of low-amplitude sound waves unless one were able to use an unrealistically high resolution. Moreover, in constructing the fluid equations we made approximations based on the assumption that all relevant length scales in the system were much larger than the mean-free-path of the molecules, but the same fluid equations produce shocks which physically have a thickness comparable to the mean-free-path, i.e. the equations have predicted their own invalidity. To allow shock handling without prohibitively high diffusivity in the entire volume, there is an additional viscosity in the vicinity of shocks. The two viscosities are one based on the sound speed and one on the divergence of the (horizontal) velocity -- the latter is very negative in a shock. They are:
\beqa\label{eq:nu}
\nu \!\!\!&=&\!\!\! \nu_1(c_{\rm s}+|{\bf u}|)\max(\Delta x,\Delta y)\\
\nu_{\rm s}\!\!\!&=&\!\!\! \nu_2\, {\rm smooth}[\max(-{\bm\nabla}\cdot{\bf u},0)]\max(\Delta x,\Delta y)^2\label{eq:nu2}
\eeqa
where smooth is a linear average defined over a cube of 3x3x3 points centred on the current one and max is defined over a cube of 5x5x5 points. Both coefficients $\nu_1$ and $\nu_2$ are dimensionless, but $\nu_2$ is generally much larger in value; despite that, the second term above is negligible except in shocks.

In addition to such a method of shock handling, many general-purpose codes use some kind of artificial diffusion scheme which can handle discontinuities and damp unwanted `zig-zags'. Here, we use a `hyper-diffusive' scheme, based on that of \citet{Nor_Gal:1995}, where the diffusion coefficients are scaled by the ratio of the third and first spatial derivatives of the quantity in question, which has the effect of increasing the diffusivity seen by structures on small scales where the third derivative is high, damping any badly resolved structure near the Nyquist spatial frequency, while allowing a low effective diffusivity on larger scales. The way this works in practice is via diffusive flux operators:
\beqa
f'_i &=& {\rm d}f_i+ \frac{\max(|{\rm d}3_{i+i}|,|{\rm d}3_i|,|{\rm d}3_{i-1}|)}{\max(|{\rm d}f_{i+i}|,|{\rm d}f_i|,|{\rm d}f_{i-1}|)} \\ 
\;\;\;\;{\rm where}\;\;\;\;\;\;\; {\rm d}f_i &=& (f_{i+\frac{1}{2}}-f_{i-\frac{1}{2}})/\Delta x\\\;\;\;\;\;{\rm and}\;\;\;\;\;\;\;\;\; {\rm d}3_i &=& {\rm d}f_{i+1} - 2{\rm d}f_i + {\rm d}f_{i-1}.
\eeqa
These flux operators replace the derivatives of quantities on which the diffusion is operating, such as inside the brackets in equation (\ref{eq:kin_diff}) below. For some kinds of diffusion we use these hyperdiffusive derivatives, and for other kinds we use the standard derivatives to give a more physical result. In addition, because of the very different length scales and grid spacings, diffusion in the vertical and horizontal directions must often be treated differently.
 
\subsubsection{Kinetic diffusion}

Assuming that bulk viscosity is zero (a good approximation in monoatomic gases), the result of viscosity is to add the following viscous force (per unit mass) to the momentum equation (\ref{eq:momentum}):
\begin{equation}\label{eq:kin_diff}
F^{\rm visc}_i = \frac{1}{\rho}\frac{\partial}{\partial x_j} \left[\rho\nu\left(\frac{\partial u_i}{\partial x_j} + \frac{\partial u_j}{\partial x_i} - \frac{2}{3}\delta_{ij}\nabla\cdot{\bf u}  \right)\right],
\end{equation}
using Einstein summation notation. Furthermore, in many applications it can be shown that the divergence term is very much smaller than the other terms, and we drop this term here, as it does not contribute in any case to numerical stability. Moreover, in many astrophysical applications including those for which this code has so far been used, the kinetic diffusivity is much smaller than the other two diffusivities and it is desirable to reduce the `effective' viscosity as much as possible whilst preserving the stability of the code. For this reason, kinetic diffusion uses hyperdiffusive derivatives (described above) inside the brackets in equation (\ref{eq:kin_diff}); the derivative outside the square brackets remains a standard high-order derivative to preserve momentum conservation. However, the hyperdiffusive derivatives are used only with the standard viscosity (equation \ref{eq:nu}) and the normal derivatives are used with the shock-viscosity (equation \ref{eq:nu2}).

The vertical and horizontal directions require different treatment. First, note that the vertical component of the viscous force is ignored, as we are not considering the vertical part of the momentum equation. Second, all terms in equation (\ref{eq:kin_diff}) which contain the vertical velocity are dropped. Third, all derivatives with respect to the vertical coordinate must be scaled by a factor $\Delta\sigma/\Delta x$ or $\Delta\sigma/\Delta y$. The viscous force (per unit mass) is:
\beqa\label{eq:viscx_sig}
F^{\rm visc}_x \!\!\!&=&\!\!\! \frac{1}{P_\ast}\left\{\frac{\partial}{\partial x}\left[P_\ast\nu\frac{\partial^*u_x}{\partial x}+P_\ast\nu_{\rm s}\frac{\partial u_x}{\partial x}\right] \right.\nonumber\\
& &\!\!\!\!\!\!\!\!\!+\frac{\partial}{\partial y}\left[\frac{P_\ast\nu}{2}\left(\frac{\partial^*u_x}{\partial y}+\frac{\partial^*u_y}{\partial x}\right)+\frac{P_\ast\nu_{\rm s}}{2}\left(\frac{\partial u_x}{\partial y}+\frac{\partial u_y}{\partial x}\right)\right] \nonumber\\
& &\!\!\!\!\!\!\!\!\!\left.+ \frac{\partial}{\partial \sigma}\left[\frac{P_\ast\nu\nu_\sigma}{2}\frac{\partial^*u_x}{\partial \sigma}+\frac{P_\ast\nu_{\rm s}\nu_\sigma}{2}\frac{\partial u_x}{\partial \sigma}\right]\right\}\\\label{eq:viscy_sig}
F^{\rm visc}_y \!\!\!&=&\!\!\! \frac{1}{P_\ast}\left\{\frac{\partial}{\partial y}\left[P_\ast\nu\frac{\partial^*u_y}{\partial y}+P_\ast\nu_{\rm s}\frac{\partial u_y}{\partial y}\right] \right.\nonumber\\
& &\!\!\!\!\!\!\!\!\!+\frac{\partial}{\partial x}\left[\frac{P_\ast\nu}{2}\left(\frac{\partial^*u_y}{\partial x}+\frac{\partial^*u_x}{\partial y}\right)+\frac{P_\ast\nu_{\rm s}}{2}\left(\frac{\partial u_y}{\partial x}+\frac{\partial u_x}{\partial y}\right)\right] \nonumber\\
& &\!\!\!\!\!\!\!\!\!\left.+ \frac{\partial}{\partial \sigma}\left[\frac{P_\ast\nu\nu_\sigma}{2}\frac{\partial^*u_y}{\partial \sigma}+\frac{P_\ast\nu_{\rm s}\nu_\sigma}{2}\frac{\partial u_y}{\partial \sigma}\right]\right\}
\eeqa
where the asterisks with the derivative signify a hyperdiffusive differentiation as detailed above, and where $\nu_\sigma$ is equal to $\Delta\sigma^2/\min(\Delta x,\Delta y)^2$.
 These values of $\nu_\sigma$ ensure not only that the units of the different terms are the same, but also that a given zig-zag structure is damped on the same number of timesteps, independently of the grid spacing in the three directions. Finally, note the role of $P_\ast$, the pseudo-density\footnote{The right term should be $P_\ast/g$, but $g$, being a constant, simplifies out of the equations.} -- its presence in this way in the equations ensures conservation of momentum.

In addition to the viscous stress, viscosity heats the fluid. The magnitude of this heating (per unit mass) is
\beq
\label{eq:ad_visc}
Q_{\rm visc} = \frac{1}{P_\ast}S_{ij}\frac{\partial u_i}{\partial x_j}
\eeq
where $S_{ij}$ is the viscous stress tensor, i.e. the contents of the square brackets in equations (\ref{eq:viscx_sig}) to (\ref{eq:viscy_sig}). The index $i$ is equal to $x$ and $y$, but the index $j$ is $x$, $y$ and $\sigma$. Units are taken care of by the presence of $\nu_\sigma$ inside the $S_{x\sigma}$ and $S_{y\sigma}$ parts of the stress tensor.

\subsubsection{Thermal diffusion}
\label{sec:therdiff}
The code includes two kinds of thermal diffusion: hyperdiffusive (similar to the momentum diffusion), and `physical', both of which are simply added to the heating per unit mass $Q$. The hyperdiffusive thermal diffusion is
\beq\label{eq:ad_therm}
Q_{\rm therm} = \frac{1}{P_\ast}{\bm\nabla} \cdot \left[P_\ast  c_{\rm p} (\kappa {\bm\nabla}^\ast T+\kappa_{\rm s} {\bm\nabla}^\ast T)\right]
\eeq
where the asterisk signifies a hyperdiffusive derivative as described above. The vertical derivatives are scaled with $\Delta\sigma^2/\Delta x^2$ as before. The diffusivities $\kappa$ and $\kappa_{\rm s}$ are calculated simply by multiplying $\nu$ and $\nu_{\rm s}$ by a number, normally unity.

It is sometimes desirable to have a larger thermal diffusion to model an actual physical process. To this end, the code also includes a physical thermal diffusion, which is simply the same as equation (\ref{eq:ad_therm}) but with standard spatial derivatives. In the vertical direction, derivatives with respect to $\sigma$ are scaled with $\partial \sigma/\partial z$. In principle, there are also cross-terms originating from the fact that surfaces of constant $\sigma$ are not horizontal. Generally though, these terms are tiny and can be dropped. Also, the difference of length scales in the horizontal and vertical often means that the physical thermal diffusion in the horizontal direction is too small to provide numerical stability, and so a hyperdiffusive horizontal diffusion is required; likewise, in many applications the physical diffusion in the vertical direction is much larger than the minimum required for stability and the vertical hyperdiffusion can be switched off.

\subsubsection{Magnetic diffusion}\label{sec:magdiff}

Finite conductivity gives rise to an extra electric field in the induction equation (\ref{eq:induction}) since in a medium with finite conductivity, an electric field in the co-moving frame is required to drive a current according to Ohm's law ${\bf J}=\sigma {\bf E}$ where here $\sigma$ is the electrical conductivity. Remembering that ${\bf J}=(c/4\pi){\bm\nabla}\times{\bf B}$ and that the magnetic diffusivity is defined as $\eta\equiv c^2/(4\pi\sigma)$, this extra field is
\begin{equation}
{\bf E}^{\rm visc} = \frac{\eta}{c}{\bm\nabla} \times {\bf B}.
\end{equation}
The code contains a hyperdiffusive scheme rather like that described above for momentum diffusion. The electric current is calculated with the standard derivatives but the diffusivity $\eta$ is scaled. The expression for the electric field is
\beq
{\bf E}^{\rm visc} = \frac{4\pi}{c^2} \, \left\{ \eta {\bf h}({\bf J}) + \eta_{\rm s }{\bf J}  \right\}
\eeq
where the hyperdiffusive operator ${\bf h}$ is
\beq
h_x = \frac{J_x}{|J_x|}\left( \Delta y^2\left|\frac{\partial^2 J_x}{\partial y^2}\right|+ \Delta \sigma^2\left|\frac{\partial^2 J_x}{\partial \sigma^2}\right| \right)
\eeq
with corresponding values for the $y$ and $z$ components.

The value of $\eta$ is given by multiplying $\nu$ by some number, normally unity. We determine $\eta_{\rm s}$ by a similar method to that used in determining $\nu_{\rm s}$, with the difference that we use the divergence not of the velocity field ${\bf u}$ but of the part of the velocity field perpendicular to the magnetic field, ${\bf u}_\perp$.

The energy consequently lost from the electromagnetic field appears as heat, the so-called Joule heating given by (per unit mass)
\beq
Q_{\rm Joule} = \frac{1}{\rho}{\bf J}\cdot{\bf E}^{\rm visc}.
\eeq

\subsubsection{Diffusion of other variables}

In addition to horizontal velocity, temperature and magnetic field, it is often necessary in the $\sigma$-coordinate scheme to apply a hyperdiffusive scheme to the other main variable, $P_\ast$. This works in exactly the same way as for the other variables, except that being two-dimensional it is somewhat simpler. The diffusivities $\nu$ and $\nu_{\rm s}$ are first averaged over $\sigma$ and multiplied by some number, normally unity, then a term is added to the partial differential equation (\ref{eq:dPstardt}):
\beq
\frac{\partial P_\ast}{\partial t} = ....... + {\bm\nabla}\cdot\left(\bar{\nu}{\bm\nabla}^\ast P_\ast+\bar{\nu}_{\rm s}{\bm\nabla}P_\ast\right),
\eeq
where the gradients are two-dimensional. This diffusion helps to damp unwanted zig-zag behaviour where it occurs.

Any additional variables must generally also have some added diffusion: this works in exactly the same way as the hyperdiffusion on other variables. Normally the diffusivity can be lower, though, if the variable is a passive tracer with no feedback on other variables.

\subsection{Parallelisation and horizontal coordinates}\label{sec:parallel}

The code has been parallelised using OpenMP, which can be used on a shared-memory machine. Parallelisation for a distributed memory machine using MPI is planned in the medium term.

Also planned for the medium term is an extension to spherical coordinates, with a view to modelling oceans and atmospheres on stars and planets.

\section{Test cases}\label{sec:tests}

In this section we describe the numerical tests used to validate the code.
We simulate the development of a shock from a
wave, the Rossby adjustment problem and the Kelvin-Helmholtz and inverse
entropy gradient instabilities. We also test the propagation of
Alfv\'en waves and the Tayler instability for the magnetic field.

All simulations have periodic horizontal boundary conditions.  We
assign the pressure via $P_{\rm{T}}$, which is constant at all times,
and $P_\ast$ (see equations\ \ref{eq:defofsigma} and
\ref{eq:dPstardt}).  We also assign the temperature and the initial
velocity fields. When initial prescriptions are better expressed in
terms of density, we assign temperature such that the right density is
regained (see equation \ref{eq:eosalpha}).

One point has to be noted about the vertical coordinate in the
figures. Given that we use the $\sigma$-coordinate system, neither
physical height nor gravity enters the equations (see equations
\ref{eq:z} and \ref{equ:fullmomentum}); only the product of the two is
present. What is really relevant is the ratio between the scale height
and the physical height of the model. What this means is that gravity
$g$ is essentially a free scaling factor which allows us to translate
our simulations to different physical settings and the choices of
gravity made here are arbitrary. However, when magnetic fields are
added physical height becomes meaningful in its own right.

\subsection{Wave developing into a shock}
\label{sec:wave}
As a first test we start with a wave developing into a shock. These
runs were performed in 2D, $x$ and $\sigma$.  The experiment is set-up
with a uniform temperature $RT=10^6$ erg g$^{-1}$ and a perturbation in
pressure (i.e. in $P_*$) such that the resulting perturbation in the
height is sinusoidal
\begin{equation} 
\delta H / H_0 = 0.1 \cos (2 \pi x / \lambda)
\end{equation}
where we used $\lambda$ equal to the extent of the domain (1 cm).

\begin{figure}
\includegraphics[width=0.93\hsize,angle=0]{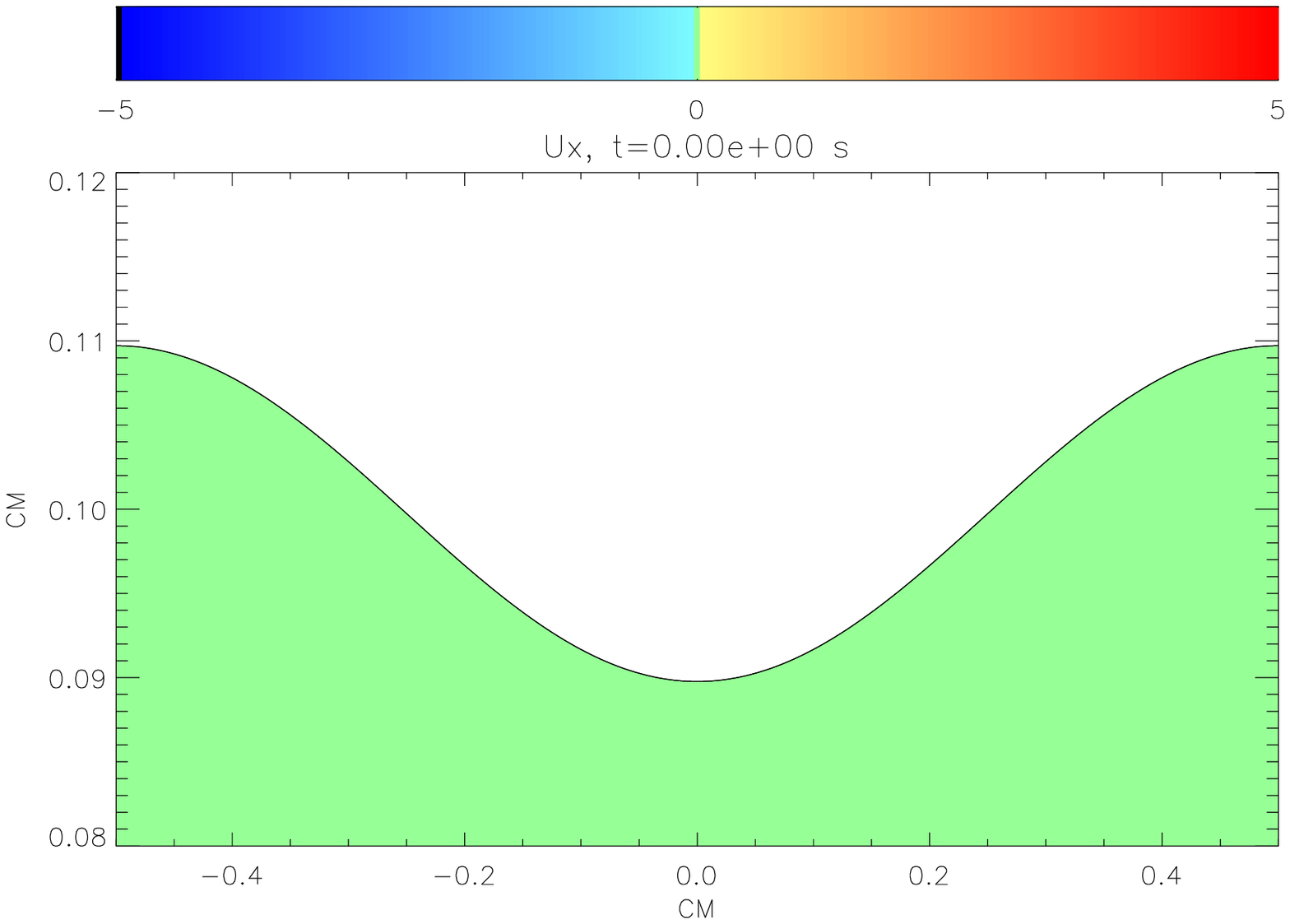}\\
\includegraphics[width=0.93\hsize,angle=0]{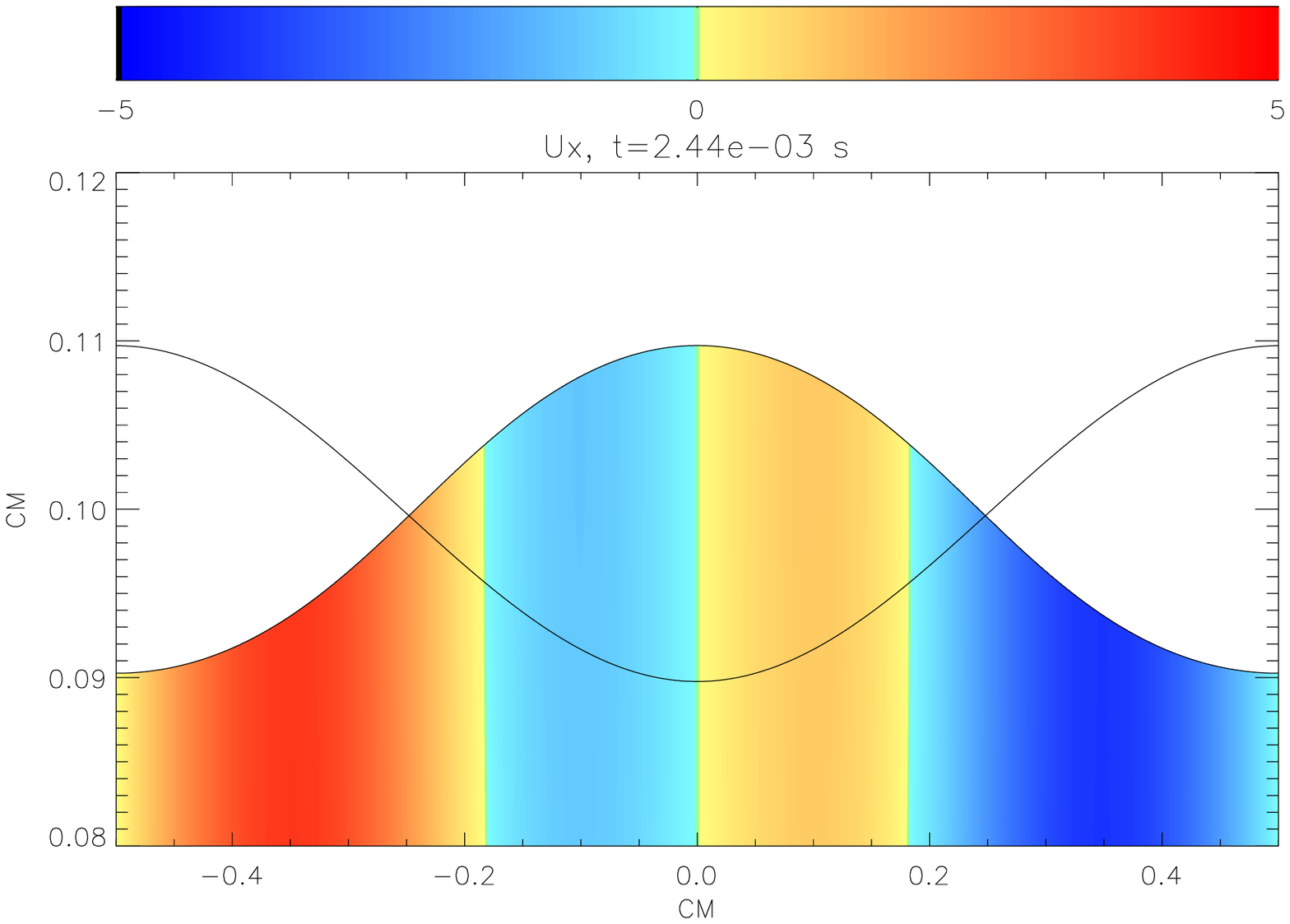}
\caption{{\it Top:} Initial snapshot of the wave simulation 200x200
  (section \ref{sec:wave}). Units of the velocity color scale in
  cm s$^{-1}$. {\it Bottom:} Snapshot at $t=2.44\times10^{-3}$ s, first
  crossing time. Units of the velocity color scale in
  cm s$^{-1}$. Superimposed is the initial height profile. The configuration
  is symmetrical with respect to the initial one.}
\label{fig:wa1}\end{figure}

\begin{figure}
\includegraphics[width=0.93\hsize,angle=0]{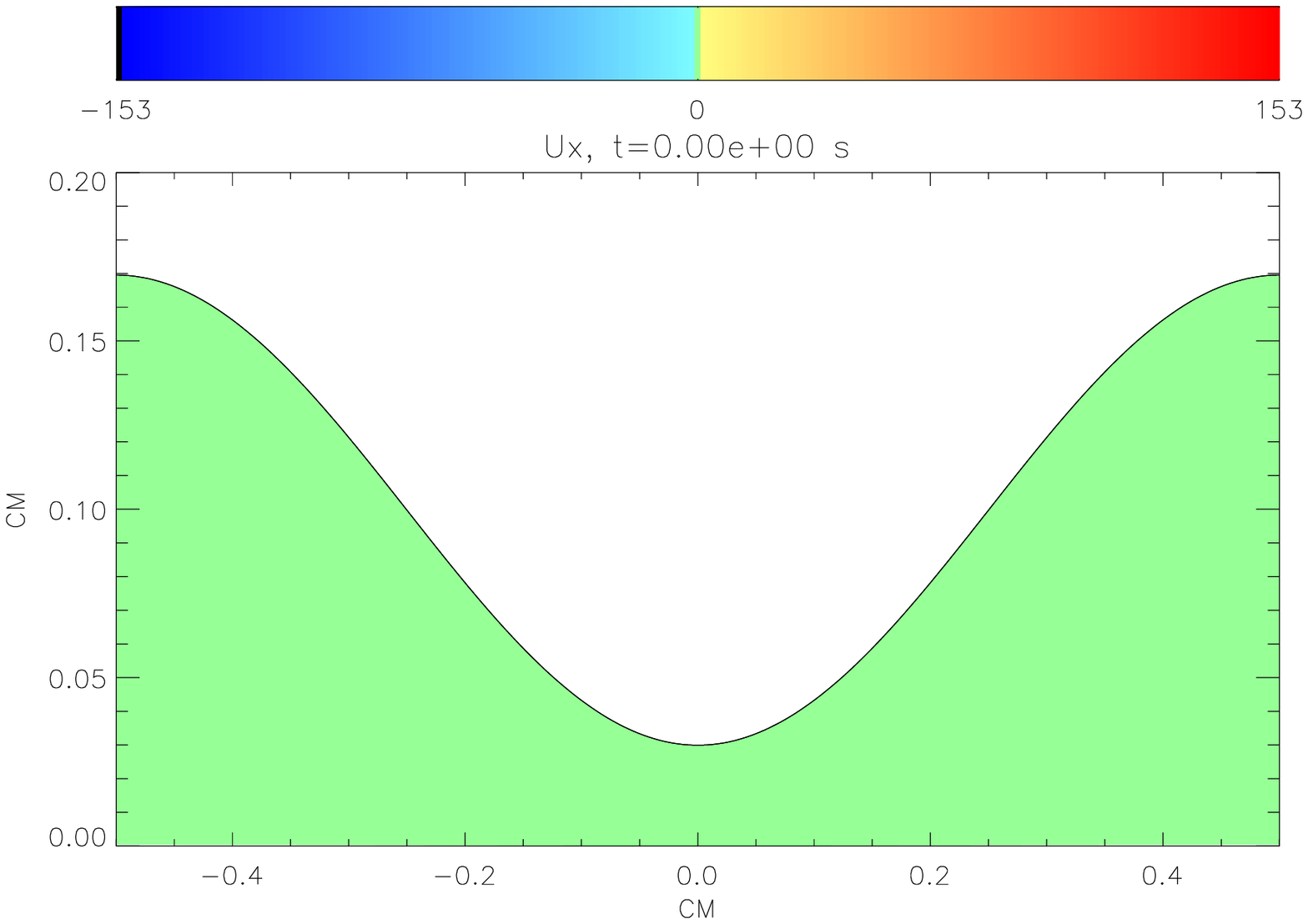}
\includegraphics[width=0.93\hsize,angle=0]{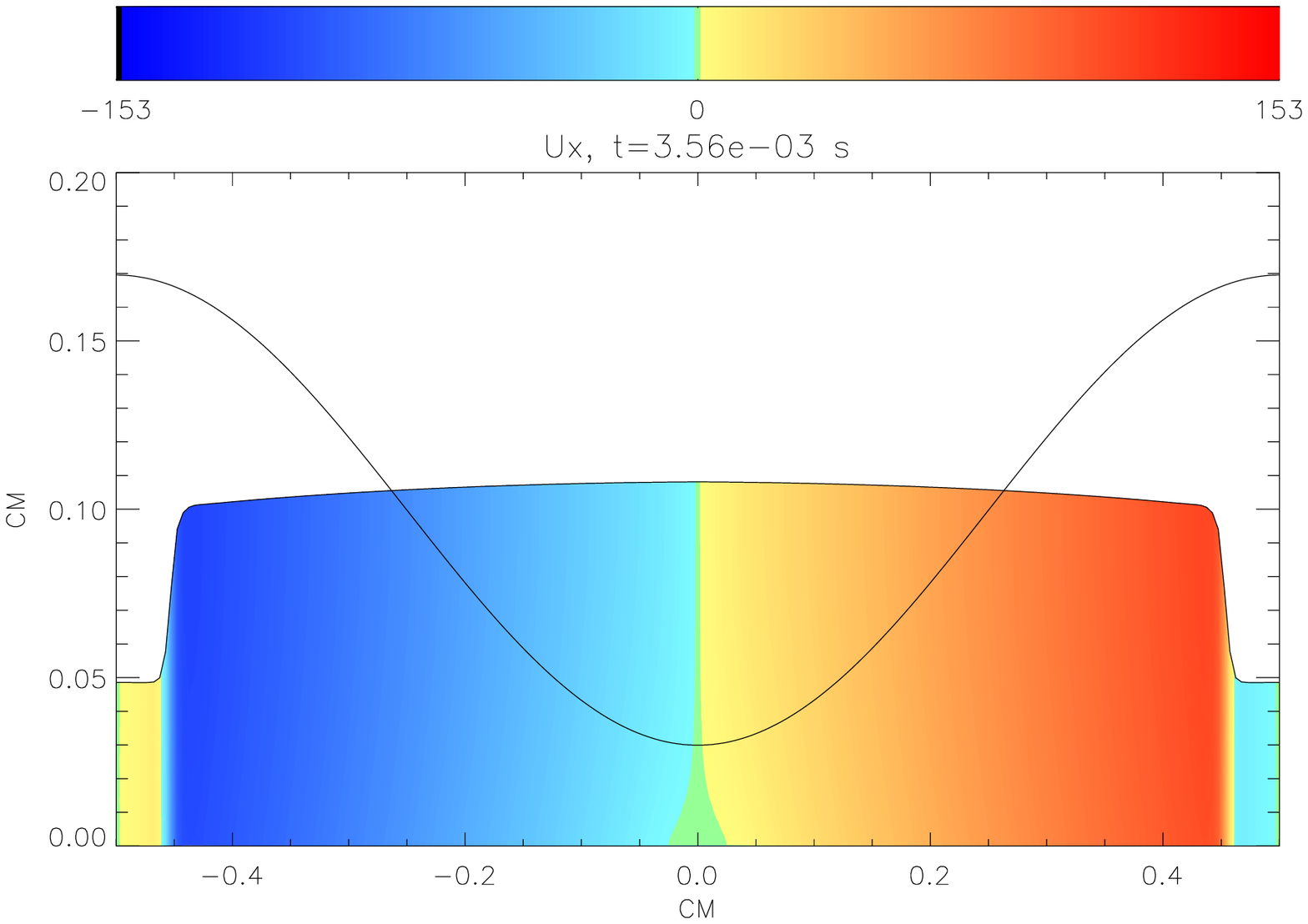}
\includegraphics[width=0.93\hsize,angle=0]{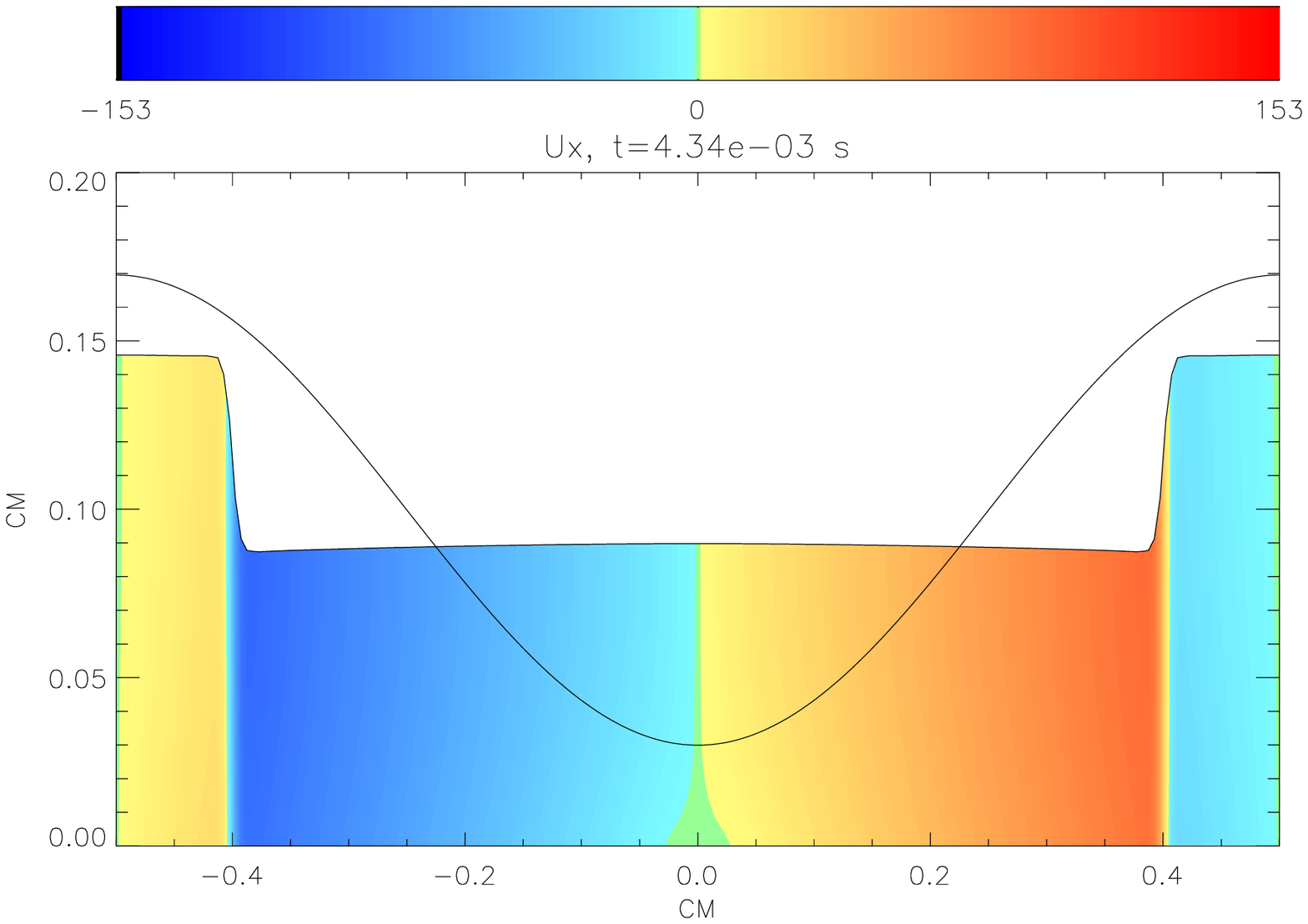}
\caption{The wave simulation 200x200, case of the strong shock
  (section \ref{sec:wave}). Units of the velocity color scale in
  cm s$^{-1}$. The three frames are a time sequence at $t=0$,
  $3.56\times10^{-3}$ and $4.34\times10^{-3}$ s. Superimposed on the
  latter two frames is the initial height profile. The shock develops
  and is then reflected off the boundary.}
\label{fig:wa3}\end{figure}

The three resolutions used to check convergence were 50x50, 100x100
and 200x200 (Fig. \ref{fig:wa1}). The wave develops into a shock
after $\sim 5$ crossing times in all three simulations. We further use
this set-up to check if the shock jump conditions are met (see below)
and to do this we also run a new simulation of a strong shock at
resolution 200x200 increasing the height perturbation amplitude to 0.7
times the background value (Fig. \ref{fig:wa3}).

\subsubsection{Conserved quantities}
\label{subsec:wavecons}

\begin{figure}
\hspace{0.3cm}\includegraphics[width=0.93\hsize,angle=0]{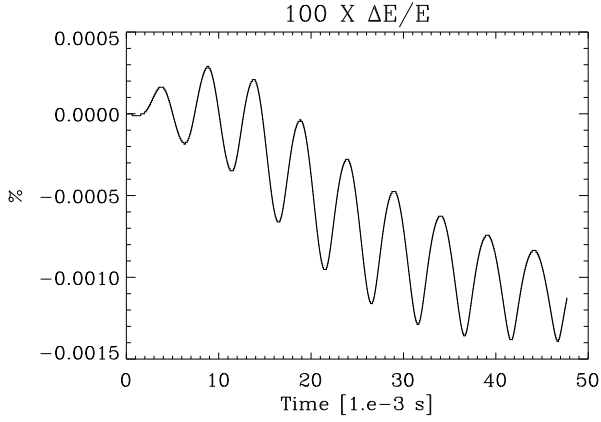}
\caption{Relative variation of $E$ for the wave simulation at
  resolution 200x200 (section \ref{subsec:wavecons}).}
\label{fig:Ewa3}
\end{figure}

At this stage it is useful to check the conservation properties of the numerical scheme, in terms of mass, momentum and energy.

We calculate the integrals as (remember that $P_\ast/g$ is ``equivalent'' to $\rho$):
\begin{align}
\label{masst}
M_{\rm{tot}}=&\int \frac{P_*}{g}\;{\rm d}x\;{\rm d}y\;{\rm d}\sigma=\frac{1}{g}\int P_*\;{\rm d}x\;{\rm d}y\\
\label{pxt}
p_{x/y,{\rm{tot}}}=&\int u_{x/y}\frac{P_*}{g}\;{\rm d}x\;{\rm d}y\;{\rm d}\sigma\\
\label{energyt}
E=&\int \left(\frac{u_x^2 + u_y^2}{2} + c_{\rm P}T\right)\frac{P_*}{g} \;{\rm d}x\;{\rm d}y\;{\rm d}\sigma
\end{align}
In the last equation \citep[see][equation 5.18]{Kasahara:74} $(u_x^2
+ u_y^2)/2$ represents the kinetic energy and $c_{\rm P} T$ is the specific enthalpy.
Note that there is no term for the gravitational potential energy -- that is because as a whole, that energy is built into the enthalpy $c_{\rm P}T$. To see this, imagine `inflating' the ocean from (close to) absolute zero whilst retaining the vertical ordering of each fluid element; for each fluid element one needs just the eventual internal energy and the $P\,{\rm d}V$ work which is used in pushing the overlying fluid upwards, and the sum of these two is simply the enthalpy; the pressure of each fluid element remains constant during this process.

We do not plot $M_{\rm{tot}}$ conservation, because it is conserved to
machine accuracy in all simulations. Fig. \ref{fig:Ewa3} shows the
evolution in time of $(E-E_0)/E_0$ for the three resolution
simulations, where $E_0$ is the initial energy. Energy is conserved to
about one part in $10^5$. In these wave simulations the conservation
of energy is essentially independent of resolution. Momentum
conservation is looked at in section \ref{sec:kh}, since here the
total momentum is zero and a fractional conservation is tricky.

\subsubsection{Wave speed}
 
We now study the velocity of the wave: considering only half the
domain, since the horizontal velocity field $u_x$ is zero at
the boundaries, before the nonlinear effects become important and the
shock develops, we may regard it as a standing wave of n=1 (the
fundamental). The velocity of the wave propagation in the medium is
\citep[see][]{Pain:05}
\begin{equation}
  c_{\rm{w}}=\omega L / \pi
\end{equation}
where $L$ is the extent of the domain (0.5 cm) and $\omega$ the wave
frequency ($2\pi/\Delta T$, $\Delta T$ its period).

The initial condition for the velocity is to be zero everywhere, we
therefore choose an arbitrary position in space ($x=0.01$ cm) and
measure the time it takes for $u_x$ to reach its minimum before rising
again. This corresponds to $\Delta T/4$. We then compute the value of
$c_{\rm{w}}$. We limit ourselves to this early stages only to avoid
nonlinear effects. The values we measure are 200.774 cm s$^{-1}$ for
the 50x50 simulation, 200.793 cm s$^{-1}$ for the 100x100 simulation
and 200.797 cm s$^{-1}$ for the 200x200 one. These are very close to
the expected value for a gravity wave with speed given by
$\sqrt{gH}=200$ cm s$^{-1}$ in shallow water approximation, although
of course we are not modelling shallow water here but shallow
\emph{compressible} gas. However at this modest amplitude the
compressibility has only a small effect.  In the case of the strong
shock we find 192.197 cm s$^{-1}$, which is less accurate, but the
perturbation is higher and the shock sets in at the first crossing,
therefore the linear approximation is definitely not valid.

\subsection{Rossby adjustment problem}
\label{sec:rossby}
\begin{figure}
\includegraphics[width=0.93\hsize,angle=0]{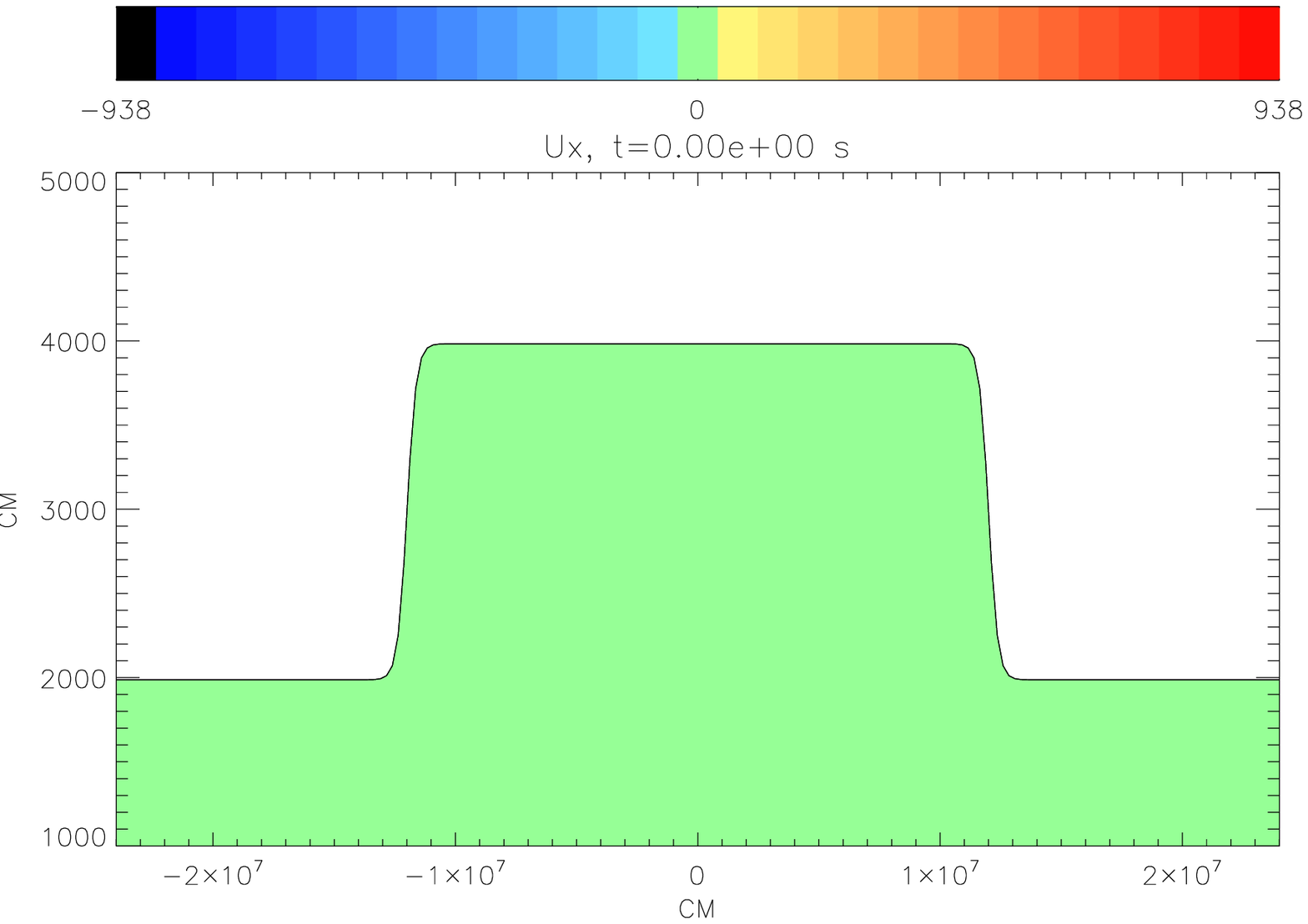}
\includegraphics[width=0.93\hsize,angle=0]{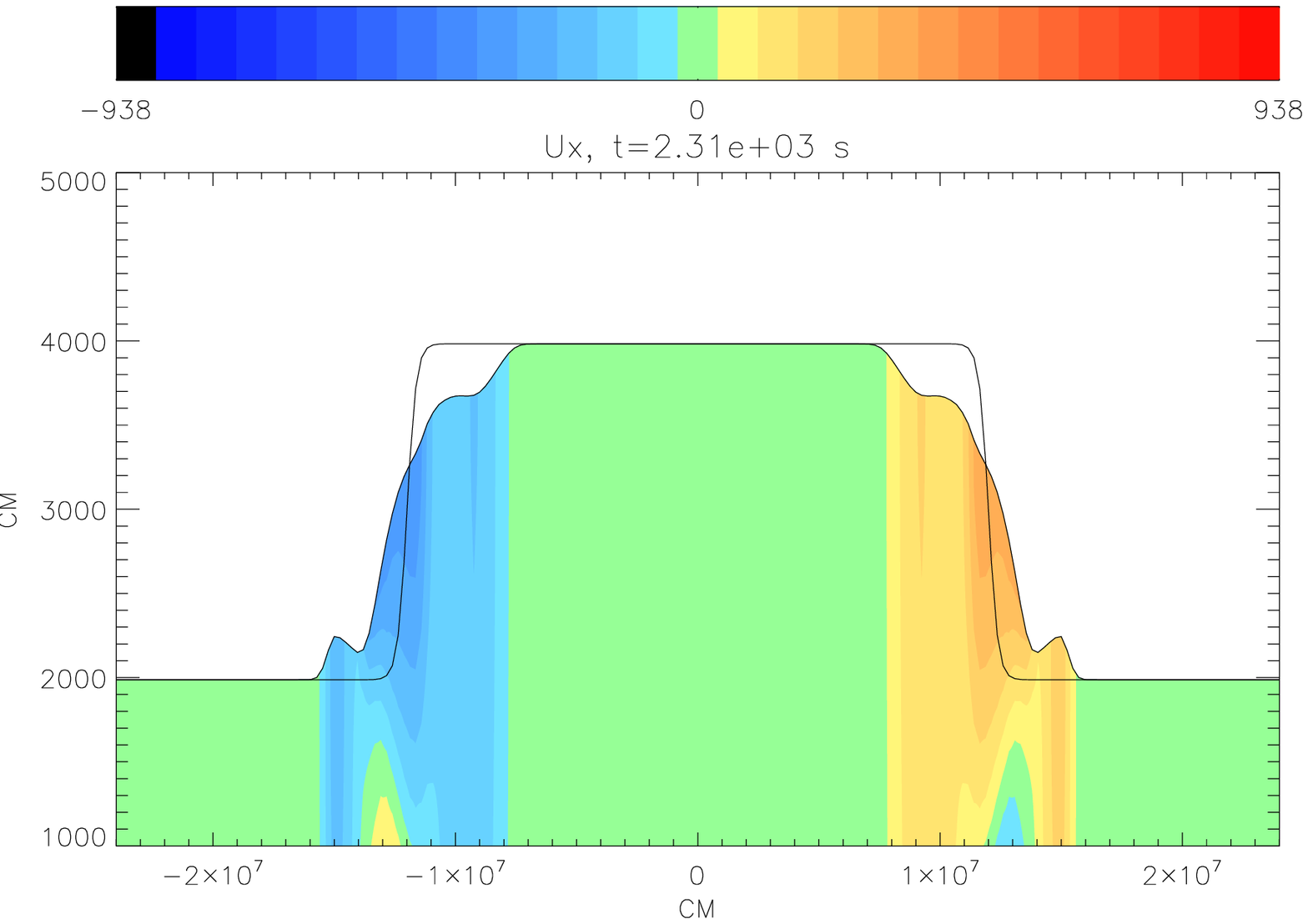}
\caption{Snapshot of the Rossby adjustment simulation 200x200 (section
  \ref{sec:rossby}) at times $t=0$ and $1.52\times10^{-3}$ s. Units
  of the velocity color scale in cm s$^{-1}$. The transient waves are
  visible.}
\label{fig:ro}
\end{figure}

We also simulate the Rossby adjustment problem. This is a 2D problem
(one vertical and one horizontal dimension) similar to the dam break,
with the addition of the effects of the rotation of the reference
frame. The fluid is assumed to be confined and at rest, until at $t=0$
s it is let free to move.  The initial conditions correspond to a
central ``bump'' in the fluid which will try to spill laterally under
the action of pressure/gravity. Coriolis force will oppose this motion
and the fluid should adjust to an equilibrium configuration with a
sloping interface that extends for $\sim 6 \rosby$ in the horizontal
($x$) direction; $\rosby$ is the Rossby radius defined as
\citep[see][]{Ped:87}:
\begin{equation}
\label{rossby}
\rosby=\frac{\sqrt{g H}}{4 \Omega}
\end{equation}
where $\Omega$ is the angular velocity of the reference frame; note that below we use the Coriolis parameter $f=2\Omega$.

The set-up for these simulations is: $g=10^3\;\rm{cm\; s}^{-2}$,
$P_{\rm{T}}=10^{4}$ erg cm$^{-3}$, $P_*=(e-1)P_{\rm{T}}$ and
$RT_{\rm{max}}=2\timt10^6$ erg g$^{-1}$. We add an initial perturbation of
the fluid temperature as
\begin{equation}
\label{rosspert}
\delta T / T_{\rm{max}} = 
\frac{ (2\eta/H_{\rm{max}} - 1) \exp{[(x-x_0)/\delta x]} }{ 1 + \exp{[(x-x_0)/\delta x}] }
\end{equation}
This perturbation goes from 0 to $2\eta/H_{\rm{max}} - 1$ over an
interval $\delta x$, being $\eta/H_{\rm{max}} - 0.5$ at $x_0$. $x_0$
is chosen to be at 75 per cent of the domain, $\delta x$ is 0.2 per
cent of it and $2\eta/H_{\rm{max}} = 0.5$. We measure the average
height $H_0$ at $x_0$.  With these choices $H_{\rm{max}}=4\times10^3$
cm, $H_0=3\times10^3$ cm and $\eta=10^3$ cm.  Again, we simulate only
half of the domain (480 km, 240 km) to save computational time and
then mirror the results and we try three resolutions of 50x50, 100x100
and 200x200 at $f=10^{-3}$ s$^{-1}$ (Fig.\ \ref{fig:ro}).

\subsubsection{Adjustment}
\label{subsec:rossbyadj}

\begin{figure*}
\includegraphics[width=0.49\hsize,angle=0]{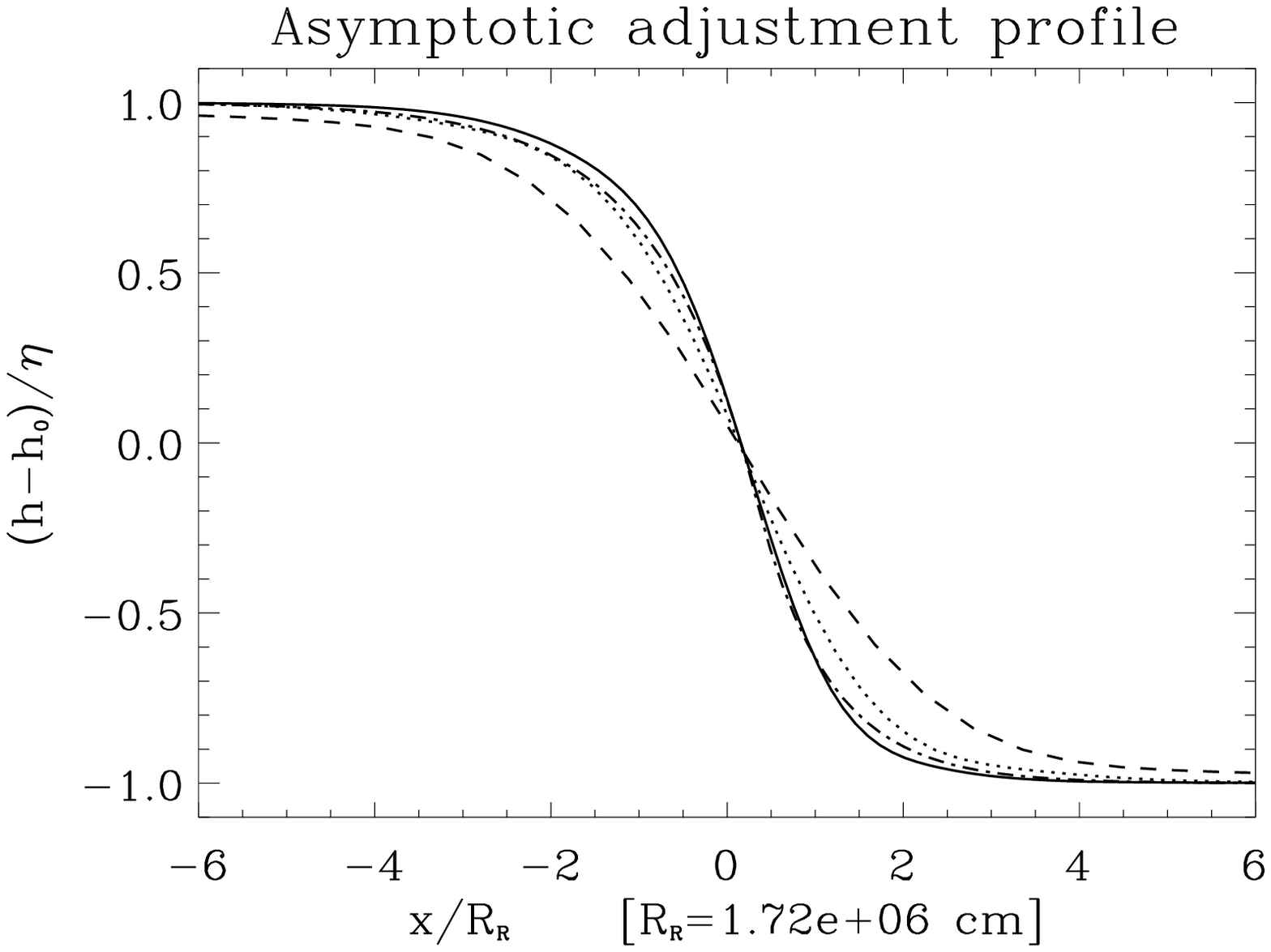}
\includegraphics[width=0.49\hsize,angle=0]{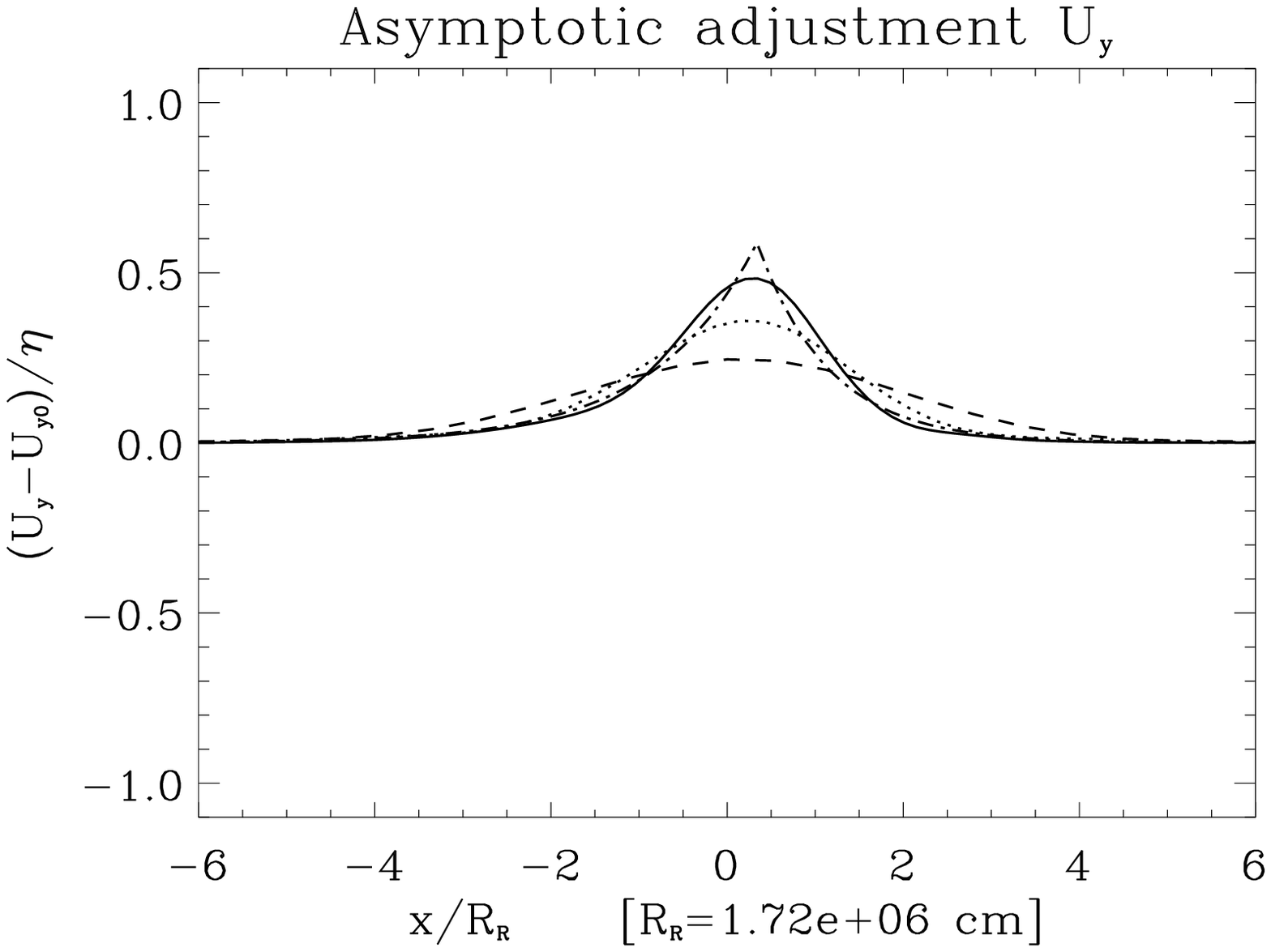}
\caption{{\it Left panel:} comparison of numerical results (for the
  Rossby adjustment simulation (section \ref{subsec:rossbyadj}) at
  resolutions 50x50 dashed line, 100x100 dotted line, and 200x200
  solid line) with theoretical prediction (equation \ref{therossh},
  dot dashed line) for the height profile. The abscissas are rescaled
  with respect to the $\rosby$ and are centred on $x_0$, see
  equation \ref{rosspert}, where the initial configuration height was
  $H_0$. {\it Right panel:} For the same simulations, a comparison of
  numerical results (dashed, dotted and solid lines) with theoretical
  prediction (equation \ref{therossvy}, dot dashed line) for the
  $u_{\rm{y}}$ profile.}
\label{fig:fronrob}
\end{figure*}

Our simulations never reached the steady state, which was to be
expected given the reflecting boundary conditions and the fact
that the time to relax can be extremely long
\citep[see][]{kuop:97}. In order to have a measure of the asymptotic
configuration we average the profile of the simulations after
$t=5\times10^3$ s when only steady gravity waves are left. The
theoretical prediction for the asymptotic shape of the profile in
shallow water approximation should be
\citep[see][]{kuop:97,both:95}:
\begin{align}
\label{therossh}
h=&\left\{\begin{array}{lr}
H_1 - \sqrt{\frac{H_1}{g}}A\exp{[\phantom{-}(x-x_{\rm{a}})/R_1]}&x\le x_{\rm{a}}\\
H_2 - \sqrt{\frac{H_2}{g}}A\exp{[          -(x-x_{\rm{a}})/R_2]}&x\ge x_{\rm{a}}\end{array}\right.\\
\label{therossvy}
u_y=&\left\{\begin{array}{lr}
A\exp{[\phantom{-}(x-x_{\rm{a}})/R_1]}&x\le x_{\rm{a}}\\
A\exp{[          -(x-x_{\rm{a}})/R_2]}&x\ge x_{\rm{a}}\end{array}\right.\\
\label{therossvz}
u_x=&0
\end{align}
but note of course that this is not completely applicable here as our
gas is compressible. $H_1=H_0 + \eta$ and $H_2=H_0 - \eta$ are the
maximal and minimal initial heights, $R_1=\sqrt{gH_1}/f$,
$R_2=\sqrt{gH_2}/f$, the Rossby radii of the two heights, and
$x_{\rm{a}}=R_1 - R_2$, $A=f x_{\rm{a}}$.

As it can be seen from Fig. \ref{fig:fronrob}, the approximation of
the theoretical results improves quite well with the resolution. This
is to be expected, since the relevant length scale $\rosby$
corresponds to only $\sim 1.8$ grid cells in the 50x50 simulation,
while it improves to $\sim 3.6$ in the 100x100 one and to $7.2$ in the
200x200 one. We measure the root mean square difference\footnote{We do
  not use the \emph{relative} difference to avoid divergences when the
  theoretical value is 0.}  between the theoretical prediction for the
profile and the numerical results: it is $1.20\times10^{-1}$,
$3.92\times10^{-2}$, and $2.22\timt10^{-2}$: definitely improving.
Also the approximation of the value of $x_{\rm{a}}$ increases:
\emph{relative} accuracy is 63.88 per cent (50x50), 60.30 per cent
(100x100), and 53.50 per cent (200x200). Fig. \ref{fig:fronrob}
(right-hand side) confirms this trend: the root mean square difference
between the theoretical prediction equation \ref{therossvy} for $U_y$
and the numerical results is $6.42\times10^{-2}$, $3.68\times10^{-2}$,
and $1.99\times10^{-2}$. Note that due to the diffusive high-order
nature of the code, we cannot reproduce the sharp peak in the
theoretical prediction and this explains the higher
discrepancies. Anyway, the approximation of the maximum value improves
steadily: relative accuracies are 58.48 per cent (50x50), 39.02 per
cent (100x100) and 17.78 per cent (200x200).

\begin{figure}
\hspace{0.3cm}\includegraphics[width=0.93\hsize,angle=0]{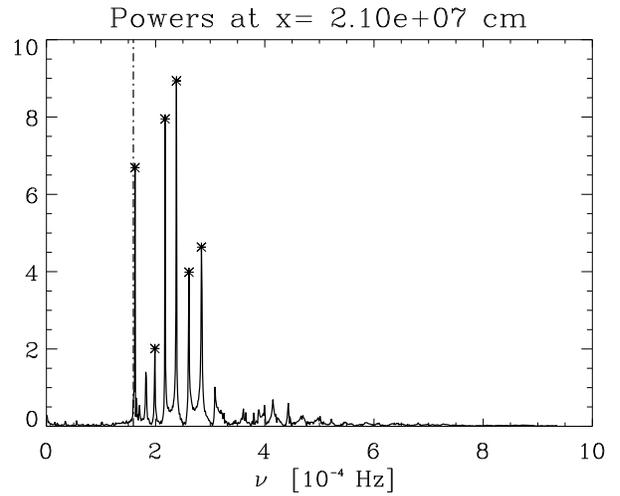}
\caption{Time Fourier analysis of the height at a fixed horizontal
  position (x=$2.1\timt10^7$ cm) for the Rossby adjustment simulation
  $f=10^{-3}$ Hz at resolution 200x200 (section
  \ref{subsec:rossbyadj}). The vertical line indicates the frequency
  corresponding to $f/2\pi$.}
\label{fig:fftrob3}
\end{figure}
\begin{figure}
\hspace{0.3cm}\includegraphics[width=0.93\hsize,angle=0]{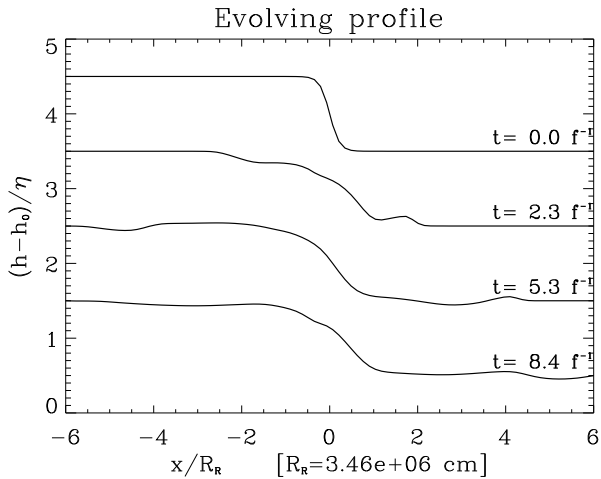}
\caption{Time evolution of the height profile for the Rossby
  adjustment simulation with $f=5\timt10^{-4}$ Hz at resolution
  200x200 (section \ref{subsec:rossbyadj}).}
\label{fig:robprof7}
\end{figure}

A Fourier analysis of the height of the surface of the gas shows the
presence of strong oscillations in addition to red noise at low
frequencies. The peaks start above $f/2\pi$, indicated as vertical
line in Fig. \ref{fig:fftrob3}, which is in
good agreement with the theoretical dispersion relation for the waves:
\citep[see][]{Ped:87}:
\begin{equation}
\nu = \sqrt{\left(\frac{f}{2\pi}\right)^2 + \frac{gH}{\lambda^2}}
\end{equation}
where $\lambda$ is the wavelength of the wave. 

Finally, as an example, Fig. \ref{fig:robprof7} shows the time
evolution of the height profile for the simulation at resolution
200x200.

\subsection{Kelvin-Helmholtz instability}
\label{sec:kh}
\begin{figure*}
\includegraphics[width=0.49\hsize,angle=0]{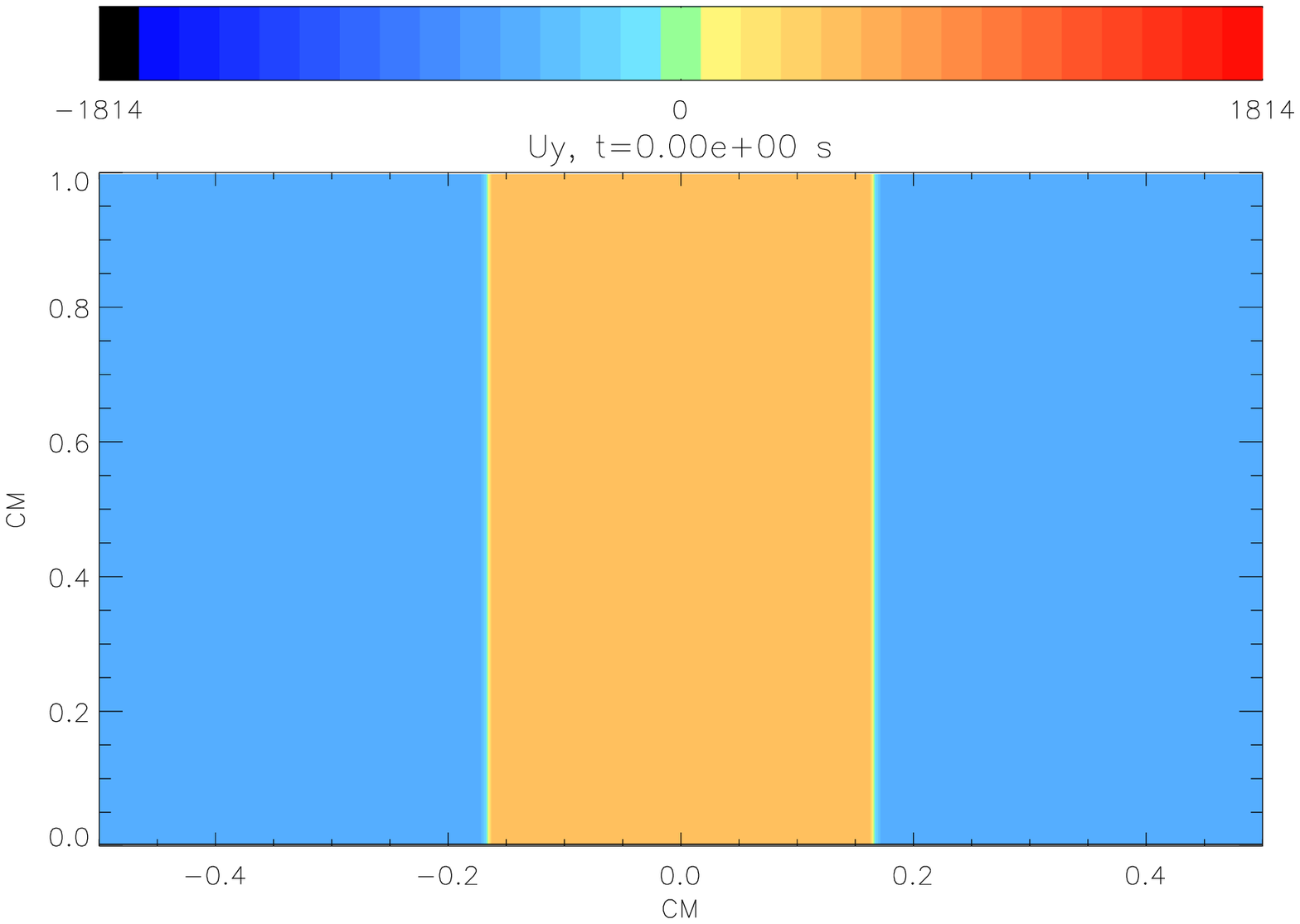}
\includegraphics[width=0.49\hsize,angle=0]{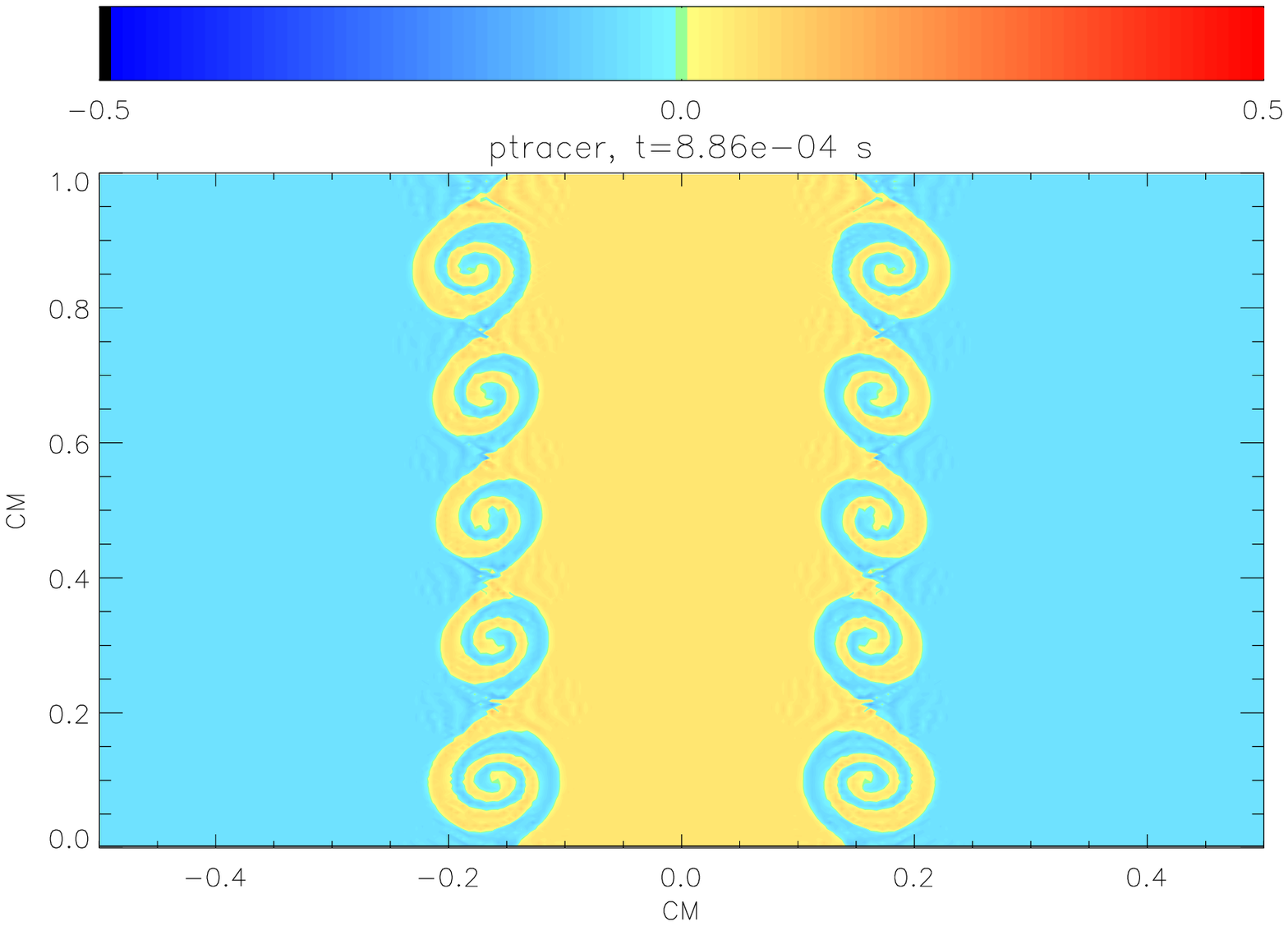}
\includegraphics[width=0.49\hsize,angle=0]{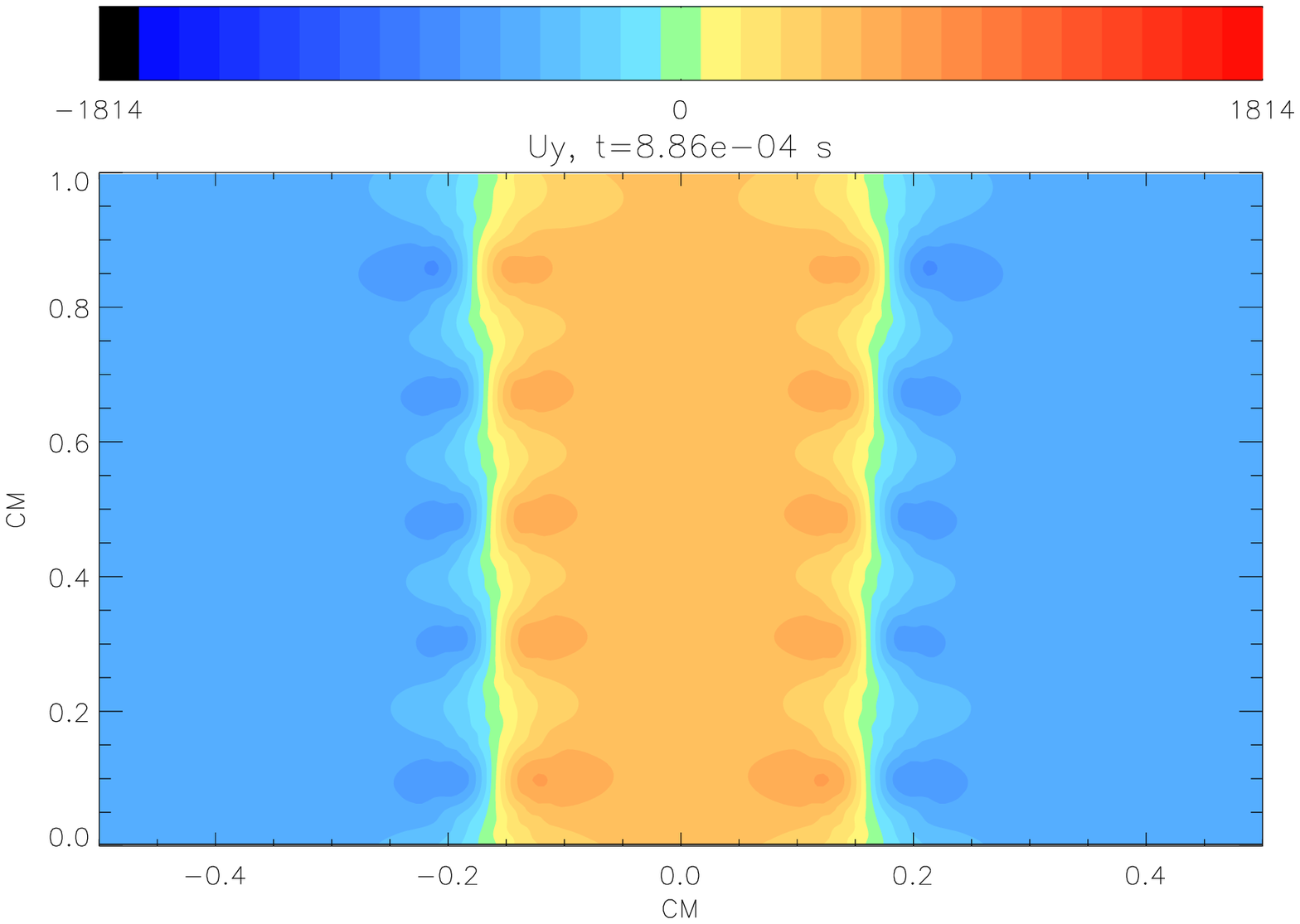}
\includegraphics[width=0.49\hsize,angle=0]{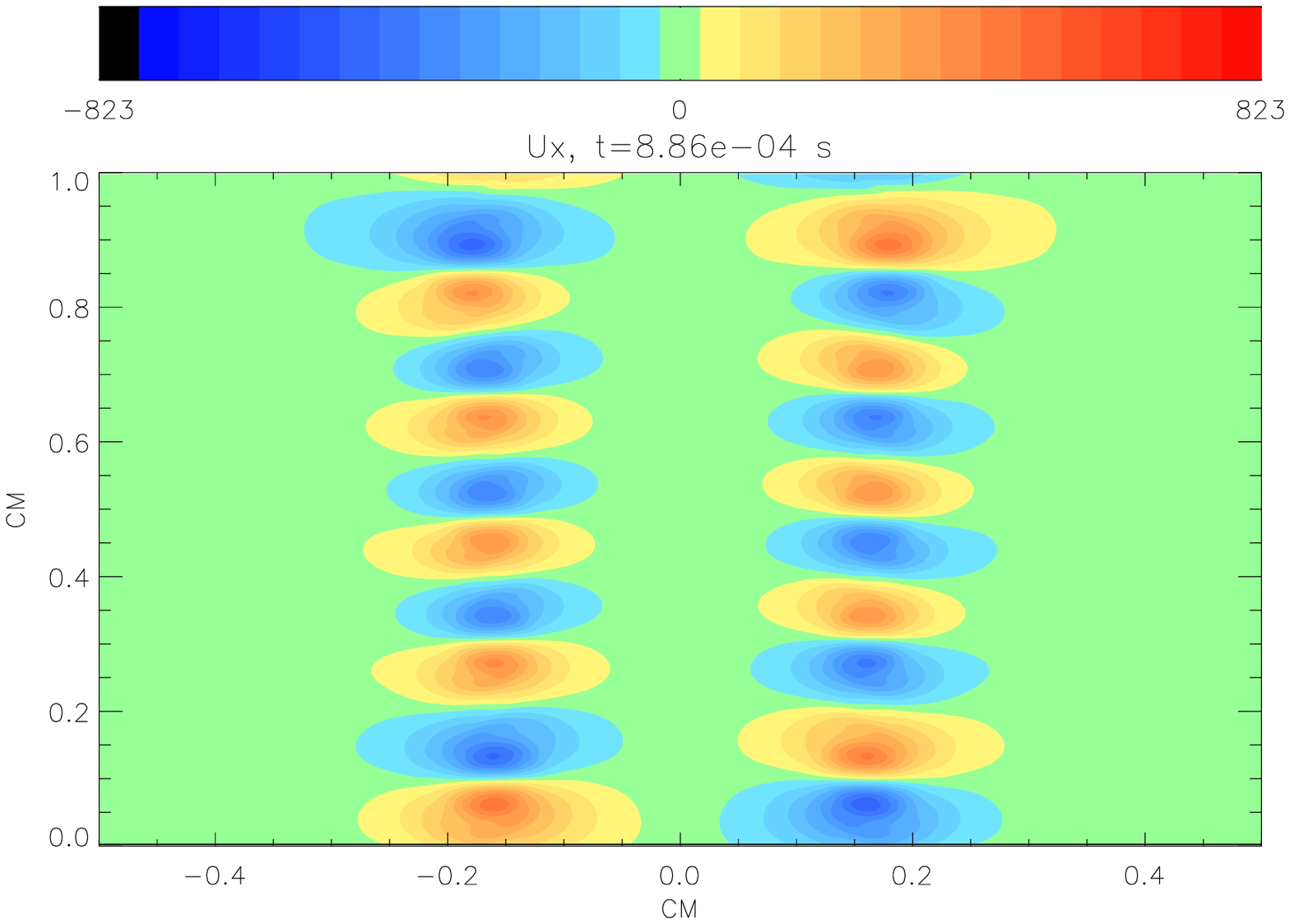}
\caption{A simulation of the Kelvin-Helmholtz instability in a square
  box with resolution 200x200x16 (section
  \ref{sec:kh}). In the initial conditions there are
  perturbations at five different wavelengths: $\lambda=1/1, 1/2, 1/3,
  1/4, 1/5$ cm. {\it Top-left:} Initial conditions, $u_y$. The
  other three frames are at $t=8.86\times10^{-4}$ s: {\it top-right:}
  passive tracer used to follow the locations of the two fluids,
  {\it bottom-left:} $u_y$, and {\it bottom-right:} $u_x$. (Note that
  these snapshots are taken at a later time than the maximum time of
  Fig. \ref{fig:khlamtot})}
\label{fig:kh}
\end{figure*}

This is a shear instability where two fluids are moving parallel to
each other with different velocities. We ran a simulation at
resolution 200x200x16 with a central section (accounting for one third
of the volume) moving with a velocity $u_y=5\times10^{3}$ cm s$^{-1}$
and the remaining two-thirds of the domain moving with equal and
opposite velocity. Initially, the temperature and $P_*$ are both
constants (sound speed $c_{\rm{s}} \sim 10^{4}$ cm s$^{-1}$), and
$P_*$ is equal to $P_{\rm T}$, so that a little under one scale height
is modelled. We follow the locations of the two fluids with the aid of
a passive tracer variable.  Finally, to get the instability started we
give the fluid an initial kick, giving the $x$-component of the
velocity with perturbations of the form $u_x=50\cos(2\pi y/\lambda)$
cm s$^{-1}$. In the following example, five wavelengths are perturbed
at once, the largest five wavelengths fitting into the domain. The
output of these simulations is plotted in Fig. \ref{fig:kh}.

\begin{figure}
\hspace{0.3cm}\includegraphics[width=0.93\hsize,angle=0]{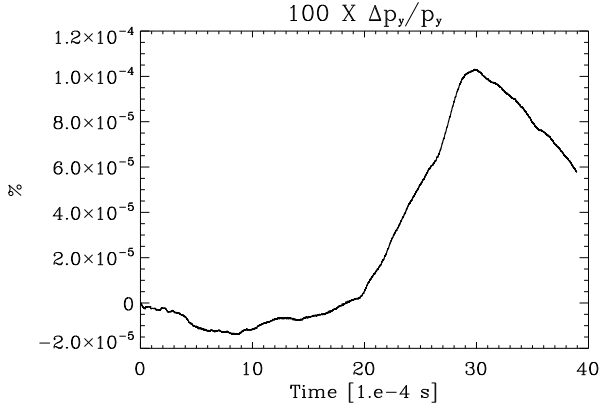}
\caption{Relative variation of $p_y$ for the Kelvin-Helmholtz
  instability simulation at resolution 200x200x16 (section
  \ref{sec:kh}).}
\label{fig:pykhb11}
\end{figure}

In this simulation, mass is conserved to machine precision as usual,
as in the simulation mentioned previously. As for momentum
conservation, which was not tested previously, the fractional change
in total momentum in the $y$-direction is plotted in Fig.
\ref{fig:pykhb11}. It is conserved here to within about one part in
$10^6$.

\begin{figure}
\includegraphics[width=0.93\hsize,angle=0]{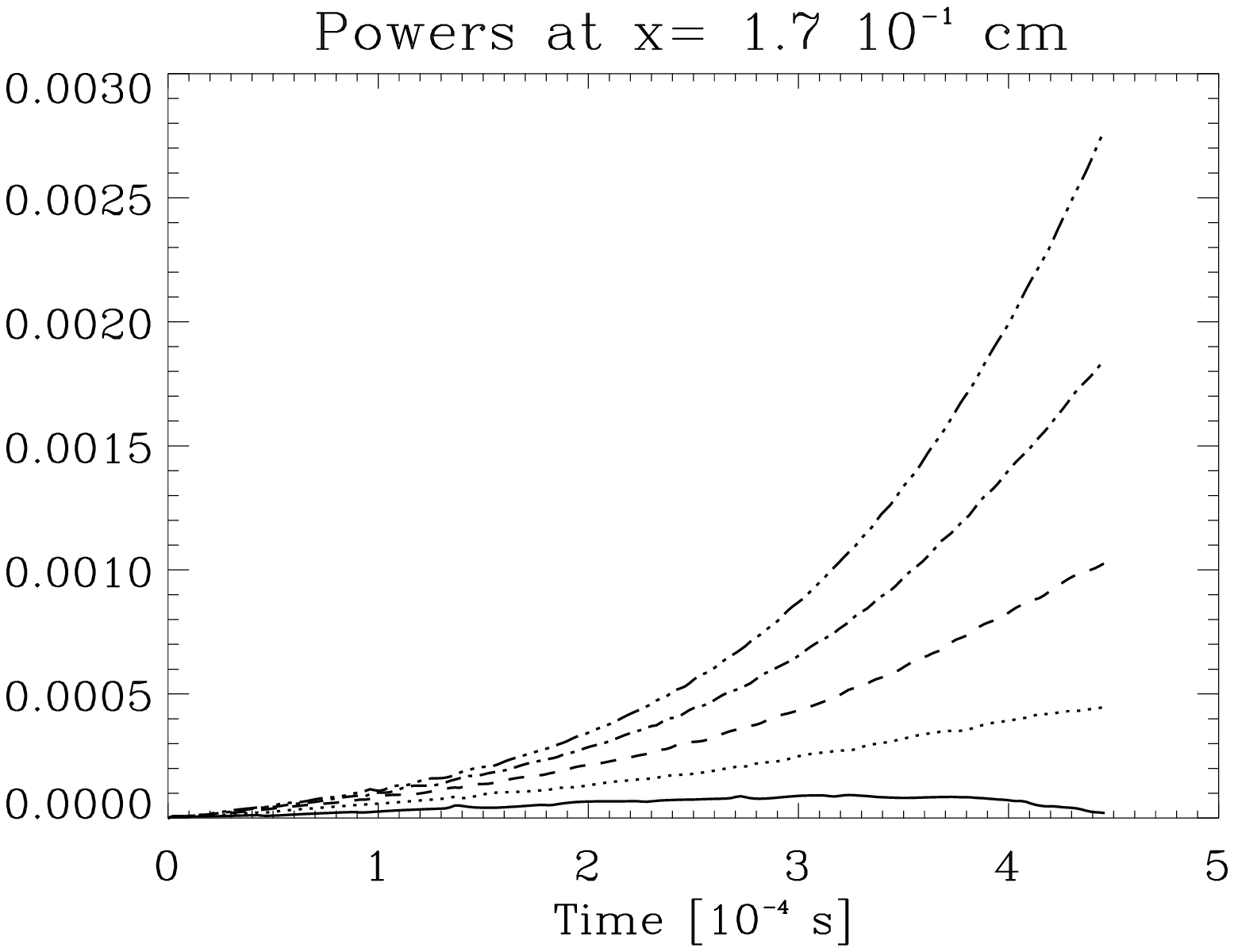}\\
\includegraphics[width=0.93\hsize,angle=0]{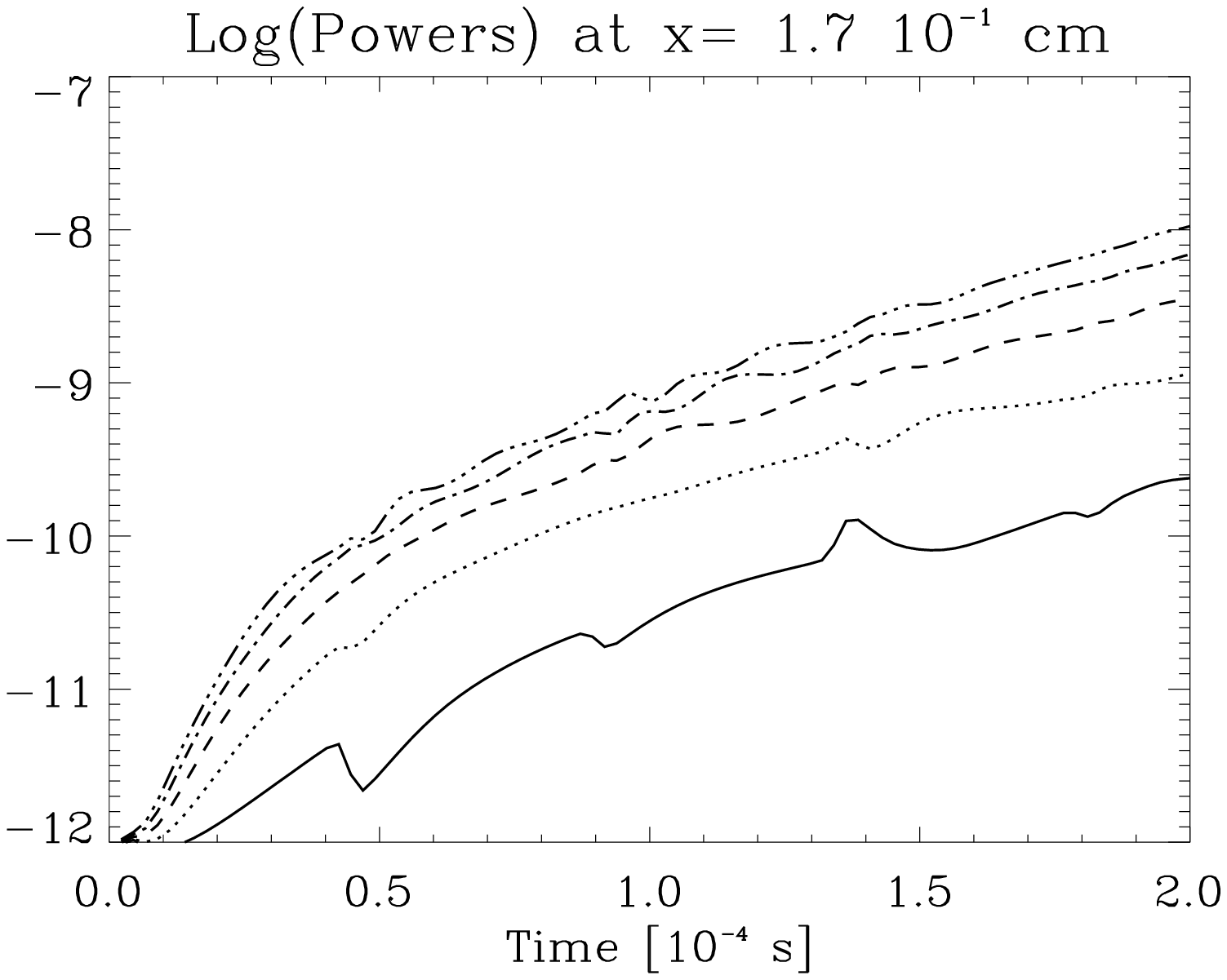}
\caption{Time evolution for the powers of $\lambda=1/1 \ldots 5$
  cm$^{-1}$ for the Kelvin-Helmholtz instability simulation at
  resolution 200x200x16 (section \ref{sec:kh}) with seeds at different
  $\lambda$s (1. solid, 1/2 dotted, 1/3 dashed, 1/4 dash dotted and
  1/5 dash dot dotted). The smallest $\lambda$ (1/5) is the first to
  grow. (Note that the snapshots of Fig. \ref{fig:kh} are taken at a
  later time)}
\label{fig:khlamtot}
\end{figure}
We now measure the growth rate of
the instability taking the Fourier transform of $u_x$ in the
$y$ direction along the line $x=1.7\timt10^{-1}$ cm (i.e. at the initial position of the interface
between the two flows). The initial perturbation should grow at a rate
$\exp{(\omega t)}$.

Under the assumptions of no stratification and incompressibility
$\omega$ is given by \citep[see][]{Chou:98}:
\begin{equation}
\label{omegakh}
\omega=\frac{2\pi}{\lambda}\sqrt{\rho_1\rho_2\left(\frac{U_{y,2}-U_{y,1}}{\rho_1+\rho_2}\right)^2}
\end{equation}
Although these assumptions are not trickily true for our simulations,
still they are good approximations and indeed looking at the amplitudes
of the five perturbed modes, we see that (as is already obvious from
Fig. \ref{fig:kh}) the shortest wavelength grows the fastest, as
predicted -- see Fig. \ref{fig:khlamtot}.

\subsection{Inverse entropy gradient instability}
\label{sec:entrop}
A common case in astrophysics is when thermal conduction is not enough
to bring heat from a lower layer to an upper one.  This situation
leads to an inverse gradient of entropy and consequently to
convection. This kind of instability can be thought of as a
\begin{figure}
\includegraphics[width=0.93\hsize,angle=0]{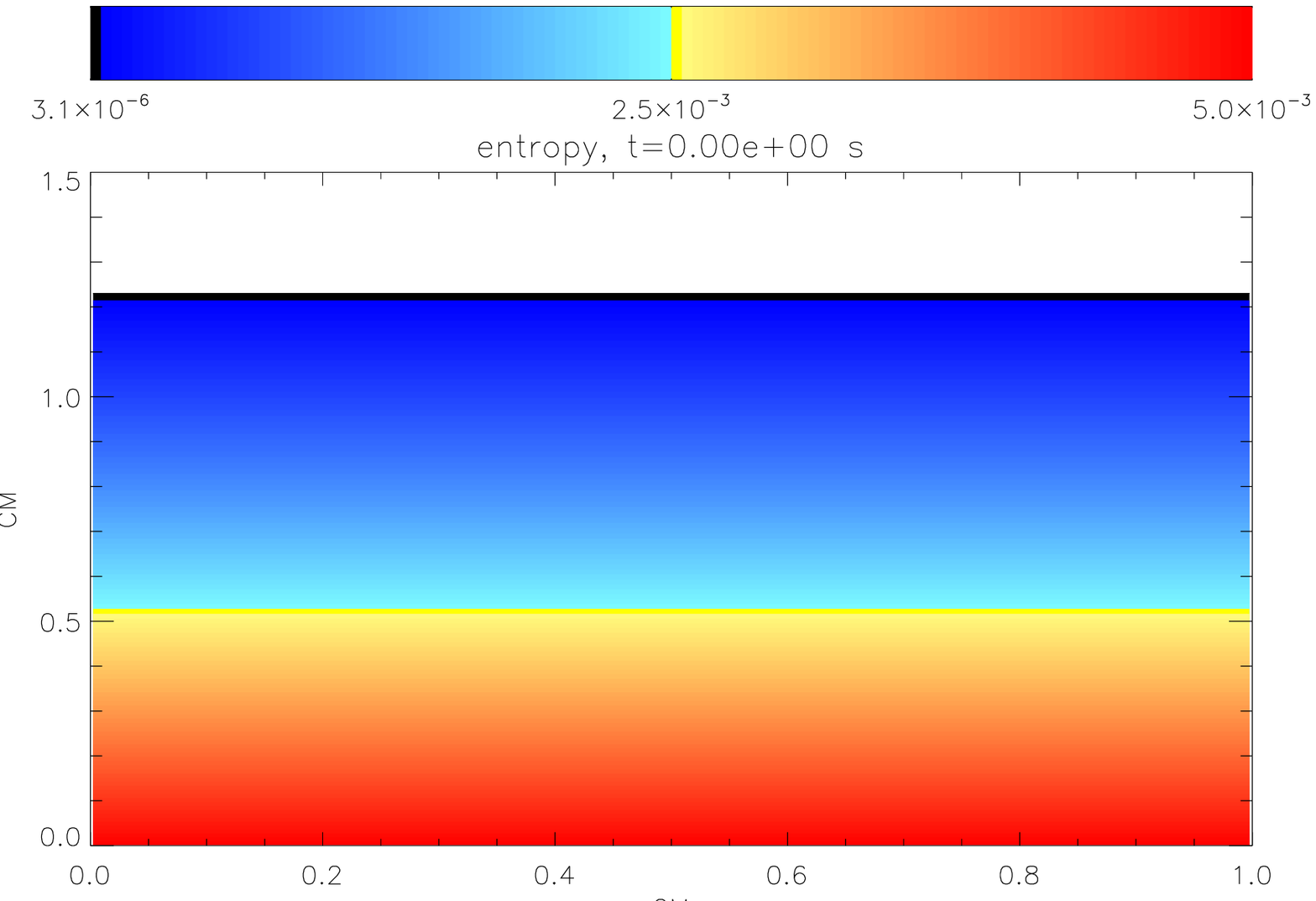}\\
\includegraphics[width=0.93\hsize,angle=0]{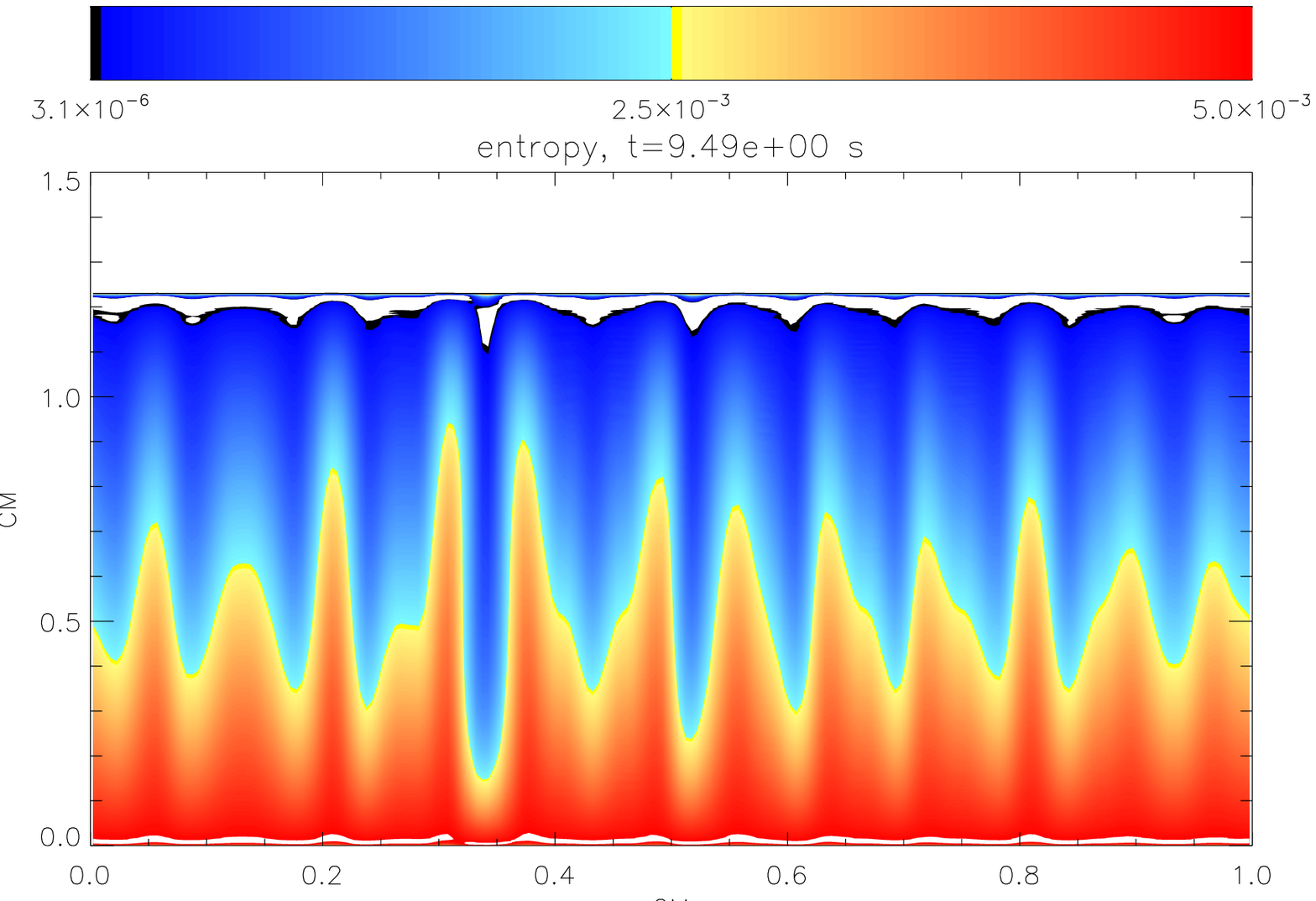}\\
\includegraphics[width=0.93\hsize,angle=0]{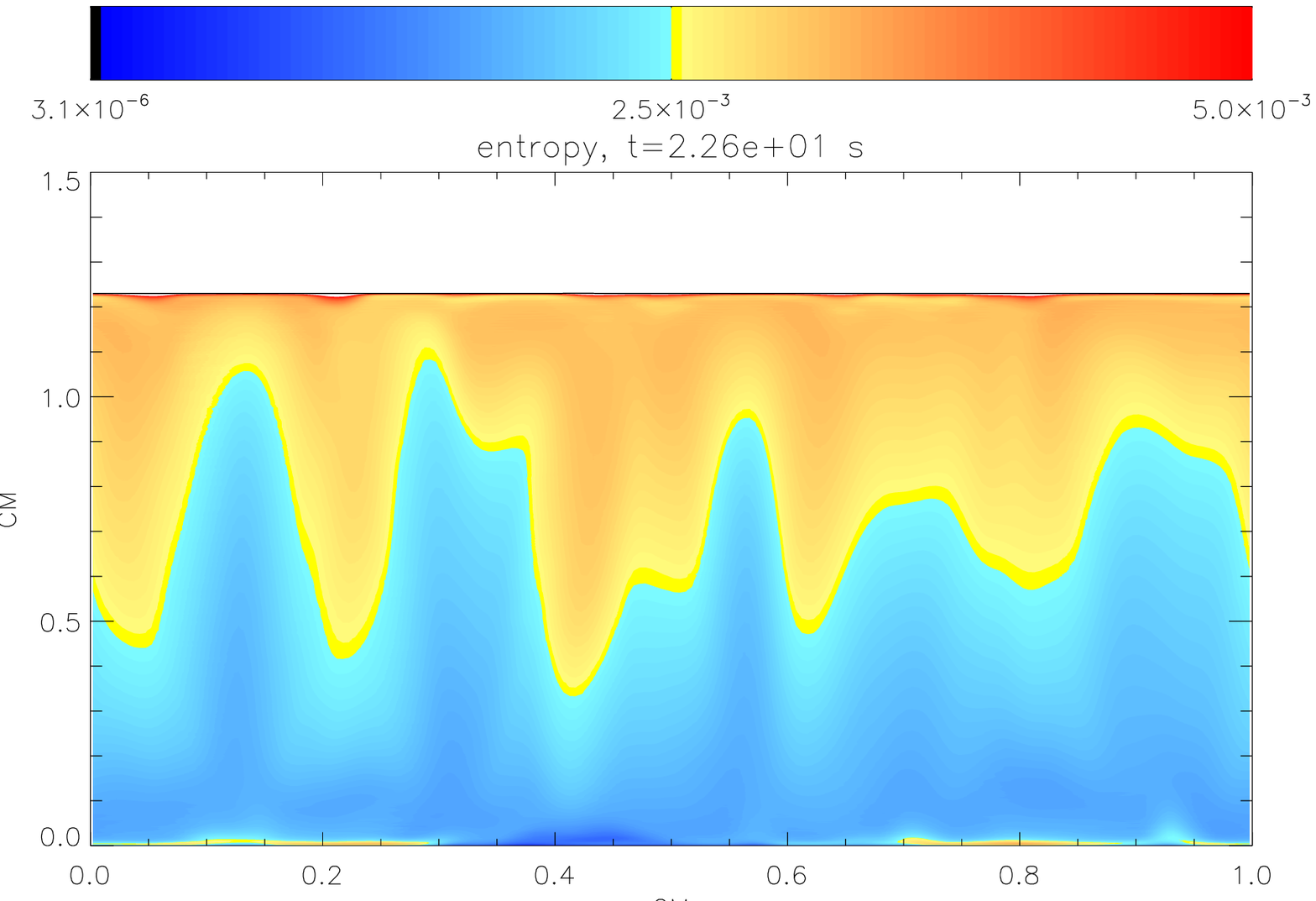}\\
\caption{Snapshots of the entropy field at $t=0,9.49,22.6$ s for the
  inverse entropy gradient simulation 200x800 (section
  \ref{sec:entrop}). Note that entropy is not fully conserved, due to
  mixing.}
\label{fig:entro}
\end{figure}
generalization of the standard text book Rayleigh-Taylor instability
and the driving force is still basically buoyancy, the main difference
being the compressibility of the gas. The criterion for stability in
case of adiabatic motion of the fluid elements is known as
Schwarzschild criterion \citep[see][]{Clay:84}:
\begin{equation}
  \label{eq:entrostab}
  {\rm d}s/{\rm d}z \ge 0
\end{equation}
where $s$ is the specific entropy and $z$ the height. In the $\sigma$-coordinate system this
translates in to
\begin{equation}
  \label{eq:entrostabsigma}
{\rm d}s/{\rm d}\sigma\le 0
\end{equation}

In the case of an ideal gas we have
\begin{equation}
 \label{eq:entro}
  s\propto \ln \left(\frac{P}{\rho^\gamma}
\right)
\end{equation}
where $\gamma=c_{\rm{P}}/c_{\rm{V}}$, which can also be rewritten, with
the use of equation \ref{eq:eos}, as:
\begin{equation}
  \label{eq:entrotemp}
  T\propto P^{1-1/\gamma}e^{s/\gamma}
\end{equation}
For our simulation, we set 
\begin{equation}
\frac{R T}{1\; \rm{erg\;g}^{-1}} = 
\left(\frac{P}{1\; \rm{erg}\;\rm{cm}^{-3}}\right)^{1-1/\gamma}
e^{0.003 \sigma}
\end{equation}
which ensures a gradient for entropy of ${\rm d}s/{\rm d}\sigma= 0.003
\gamma> 0$ in violation of condition
(\ref{eq:entrostabsigma}). $P_{\rm{T}}=1$ erg cm$^{-3}$ and
$P_\star=\left(e-1\right)P_{\rm{T}}$, so that we simulate 1 scale
height. The average sound speed is $\sim 1.5$ cm s$^{-1}$. To start the
instability we perturb the initial velocity field as:
\begin{equation}
  u_x=1\timt 10^{-6}\sum_{i=1}^{12} 
\sin \left(\frac{2\pi}{\lambda_i}x + \varphi_i\right) \rm{cm \;s}^{-1}
\end{equation}
where $\lambda_i=1/i$ cm and $\varphi_i$ is a set of random
phases. The domain extent is 1 cm.

In order to make sure that the motion of fluid elements is as adiabatic as possible
we include just the hyperdiffusive thermal conduction (see
  sec. \ref{sec:therdiff}). This test is run in 2D with a resolution of 200x800 and
Fig. \ref{fig:entro} shows the initial conditions and the evolution of
the entropy profile: after $\sim 20$ s the profile has completely
overturned.

\begin{figure}
\includegraphics[width=0.93\hsize,angle=0]{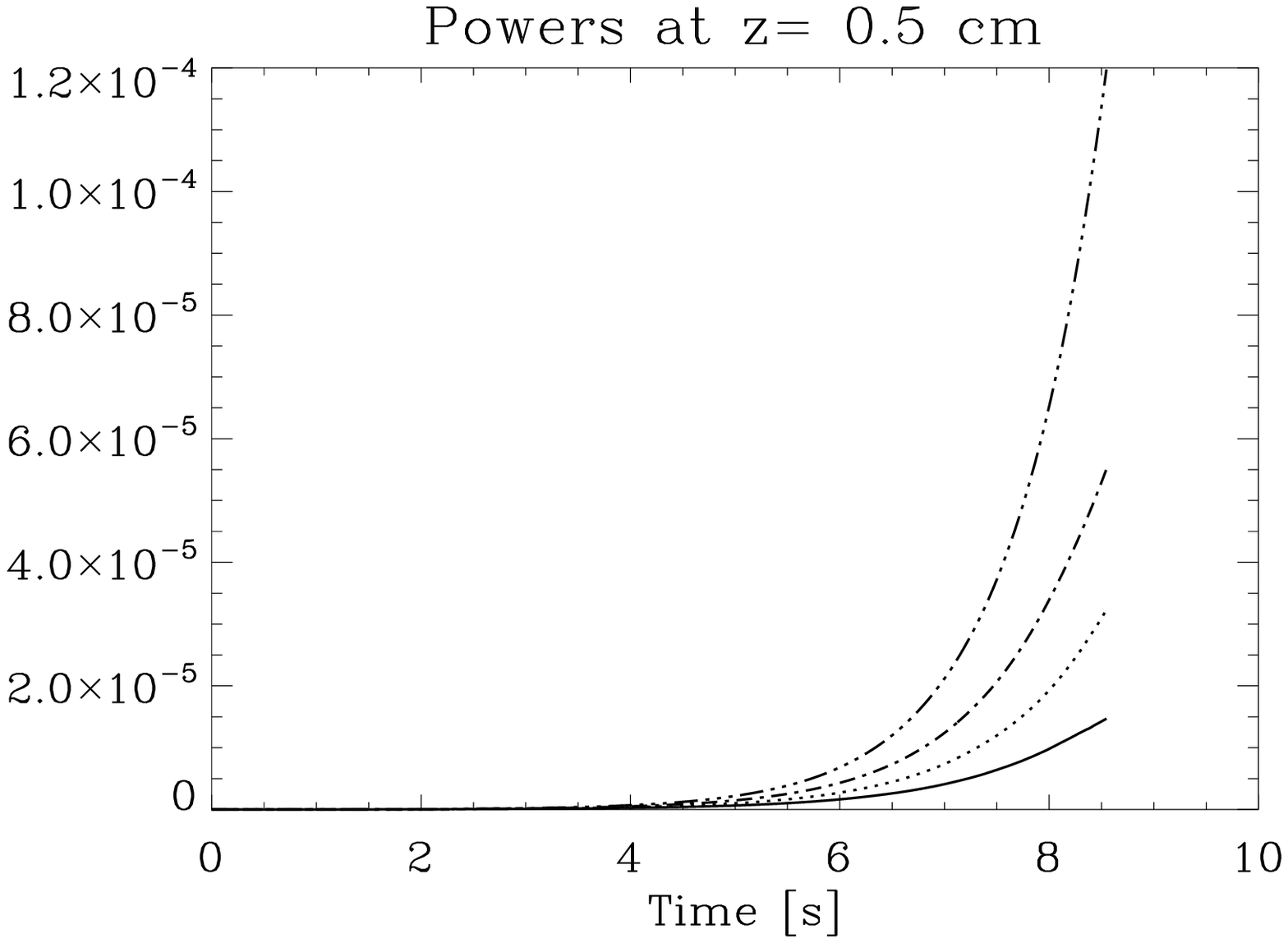}\\
\includegraphics[width=0.93\hsize,angle=0]{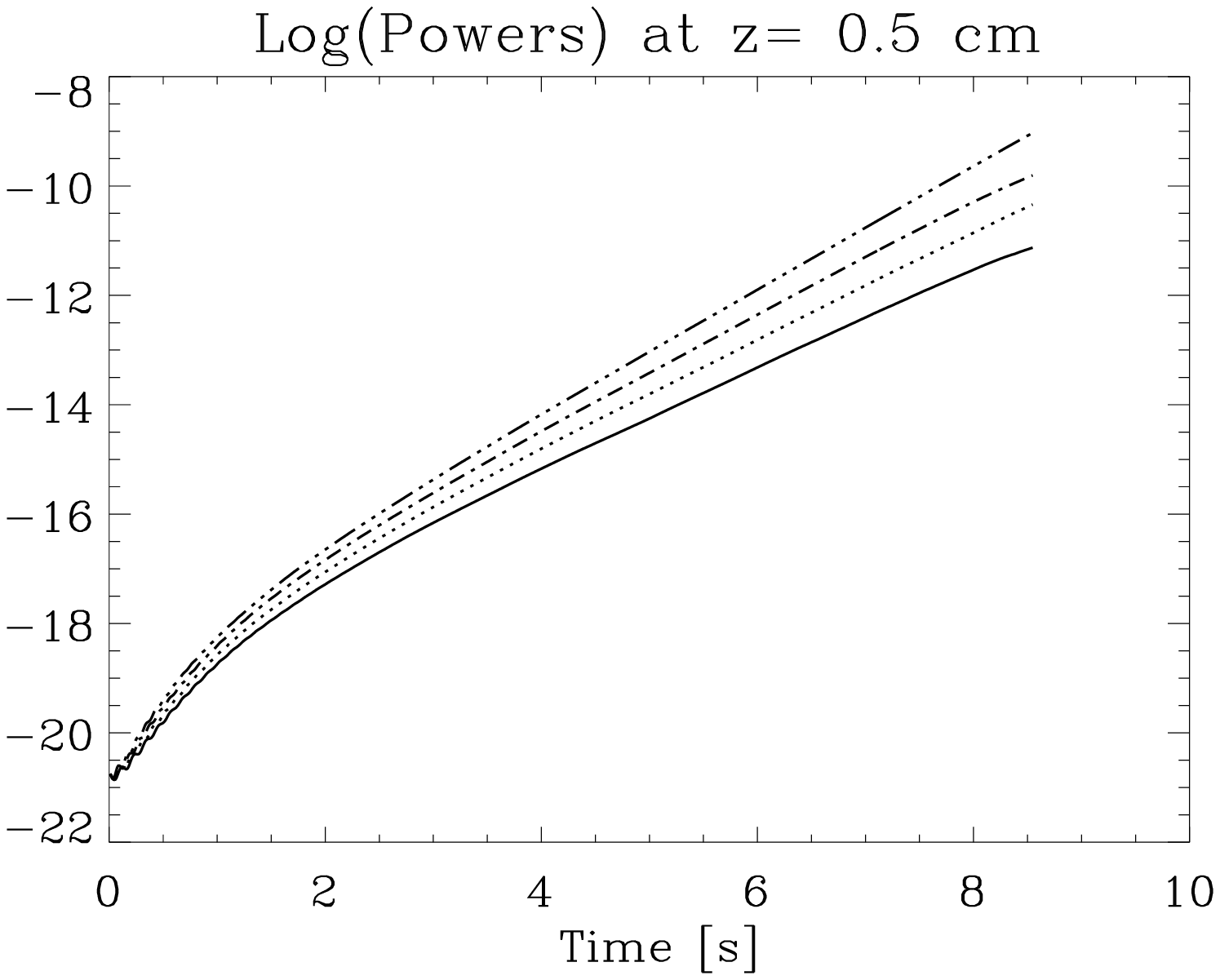}\\
\caption{Time evolution for the powers of $\lambda=1/9 .. 12$
  cm$^{-1}$ (solid, dotted, dot dashed, dot dot dot dashed) for the
  linear regime of the inverse entropy gradient simulation 200x800
  (section \ref{sec:entrop}). The growth rate is proportional to the
  wavenumber $1/\lambda$.}
\label{fig:entro-pow}
\end{figure}

The growth rate of the instabilities in the linear regime is of order
\begin{equation}
\omega \propto \frac{\sqrt{g H_{\rm{p}}\left(\nabla_{\rm{ad}} - \nabla\right)}}
{\lambda}
\end{equation}
where $\omega$ and $\lambda$ have the same meaning as in section
\ref{sec:kh}, $g$ is the gravitational acceleration, $H_{\rm{p}}$ is
the scale height. $\nabla=d \log(T) / d \log(P)$ and
$\nabla_{\rm{ad}}$ is the derivative in the adiabatic case (for a
perfect gas $\nabla_{\rm{ad}}=0.4$). Therefore, smaller wavelengths
should develope first. This is indeed the case and in figure
\ref{fig:entro-pow}
we show the time evolution of the fourier powers and logarithms of the
powers while the simulation is still in the linear regime. The growth
rate (the slope of the $\log$ plots) is  proportional
to the wavenumber $1/\lambda$.

\subsection{Magnetic field tests}

In this section the implementation of magnetic fields is tested, by propagating Alfv\'en waves and by modelling the Tayler instability in a toroidal field.

\subsubsection{Alfv\'en waves}
\label{subsec:alfv}

\begin{figure}
\hspace{0.3cm}\includegraphics[width=0.93\hsize,angle=0]{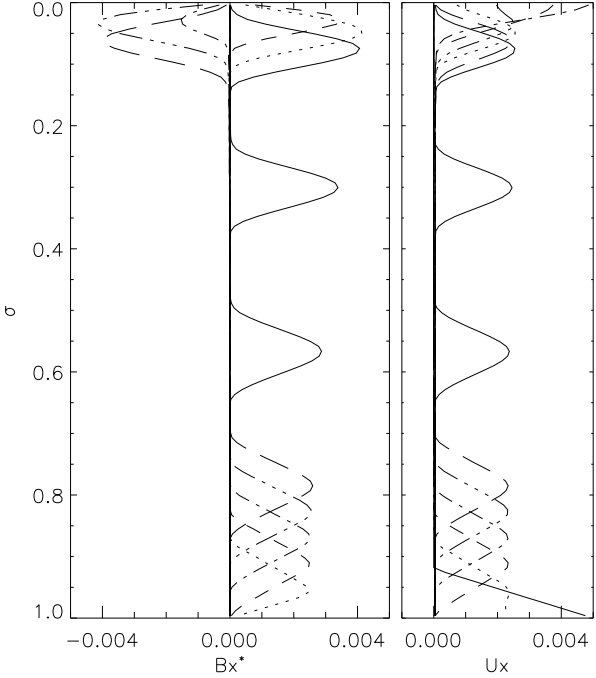}
\caption{Simulation of the propagation of a plane Alfv\'en wave in the vertical direction (section \ref{subsec:alfv}) -- a time sequence of $B_x^*$ (left) and $u_x$ (right). The first six points in time (solid, dotted, dashed, dot-dashed, dot-dot-dot-dashed, long-dashed) near the bottom of the plot are at times $t=0$, $46$, $93$, $139$, $186$ and $232$ s. The next two -- the solid lines half way up, are at times $t=487$ and $835$ s. The final six (with the same line-styles as the first six) are at times $t=1171$, $1218$, $1264$, $1310$, $1357$ and $1403$ s. Note how the amplitude of the wave increases as lower density is reached, as viewed in $B_x^*$.}
\label{fig:vert_a}
\end{figure}
\begin{figure}
\hspace{0.3cm}\includegraphics[width=0.93\hsize,angle=0]{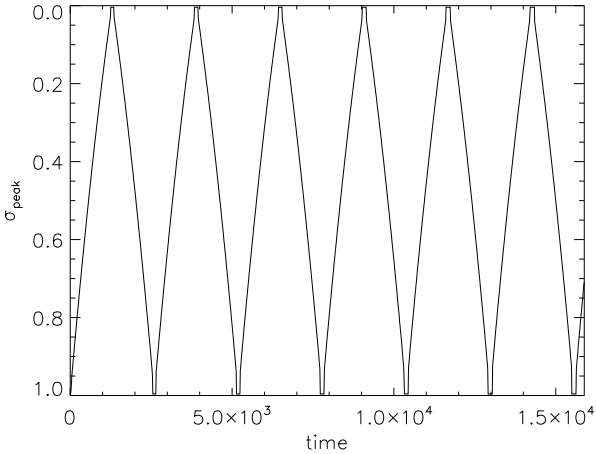}
\caption{Simulation of the propagation of a plane Alfv\'en wave in the vertical direction (section \ref{subsec:alfv}) -- the position (in terms of coordinate $\sigma$) of the peak of the wave, as it propagates back and forth between top and bottom.}
\label{fig:vert_b}
\end{figure}
\begin{figure}
\hspace{0.3cm}\includegraphics[width=0.93\hsize,angle=0]{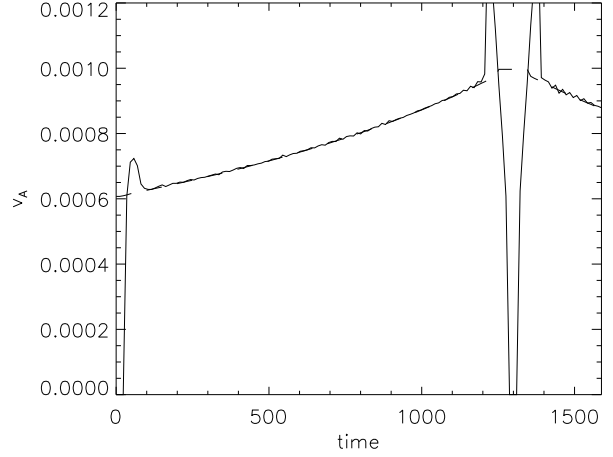}
\caption{Simulation of the propagation of a plane Alfv\'en wave in the
  vertical direction (section \ref{subsec:alfv}) -- the speed of
  propagation of the wave (solid line) compared to the theoretical
  prediction (dashed line). The two agree very closely, except when
  the wave bounces off the boundaries at $t\sim 1300$ s, when it is
  impossible to measure the propagation speed properly.}
\label{fig:vert_d}
\end{figure}
\begin{figure}
\hspace{0.3cm}\includegraphics[width=0.93\hsize,angle=0]{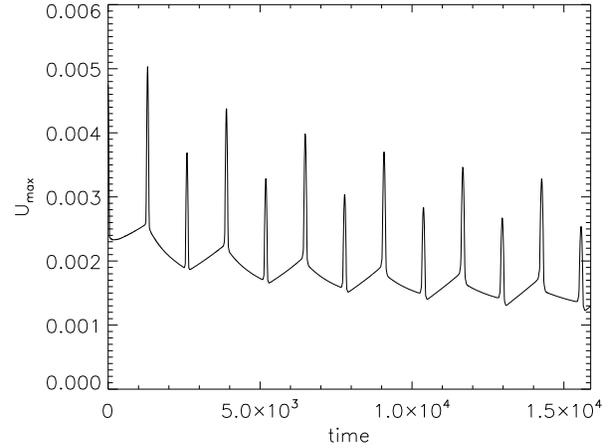}
\caption{Simulation of the propagation of a plane Alfv\'en wave in the vertical direction (section \ref{subsec:alfv}) -- the amplitude of the wave (as measured by peak $u_x$) against time, showing a gradual decay. This decay is reduced at higher resolution and/or smaller diffusion coefficients.}
\label{fig:vert_c}
\end{figure}

We test here the propagation of a plane Alfv\'en wave in the vertical
direction. The initial magnetic field is simply a uniform field
$B_z=B_0$, and it is set in motion with an initial velocity field
$u_x=u_0\max(0,(\sigma-\sigma_0)/(1-\sigma_0))$, which is just a
`hockey-stick' shape with non-zero value at $\sigma$ between
$\sigma_0$ and 1. We set $\sigma_0=0.92$, so we have a kick at the
bottom of the domain. Periodic boundaries are used in the two
horizontal directions; in the vertical, we use antisymmetric
conditions for $B_\parallel$ and symmetric for $B_\perp$, which, as we
said, are similar to the ``pseudo-vacuum'' boundaries. The
computational domain has a height equal to one scale-height, and the
temperature is uniform. As can be seen in Figs. \ref{fig:vert_a} to
\ref{fig:vert_c}, the wave propagates upwards, growing in amplitude as
it does so in response to the lower density higher up, reflects from
the upper boundary and propagates back downwards. We follow the
propagation through a number of journeys between top and bottom,
finding that the wave is damped only rather slowly. The speed of
propagation of the wave is compared to the local Alfv\'en speed
\citep{Chou:98} $v_{\rm{A}}=B_0/\sqrt{4\pi\rho}$ in Fig.
\ref{fig:vert_d}, where we can see that the agreement is close. The
vertical flux is conserved perfectly in this simple setup.

\subsubsection{Tayler instability}
\label{subsec:tayler}

\begin{figure}
\hspace{0.3cm}\includegraphics[width=0.93\hsize,angle=0]{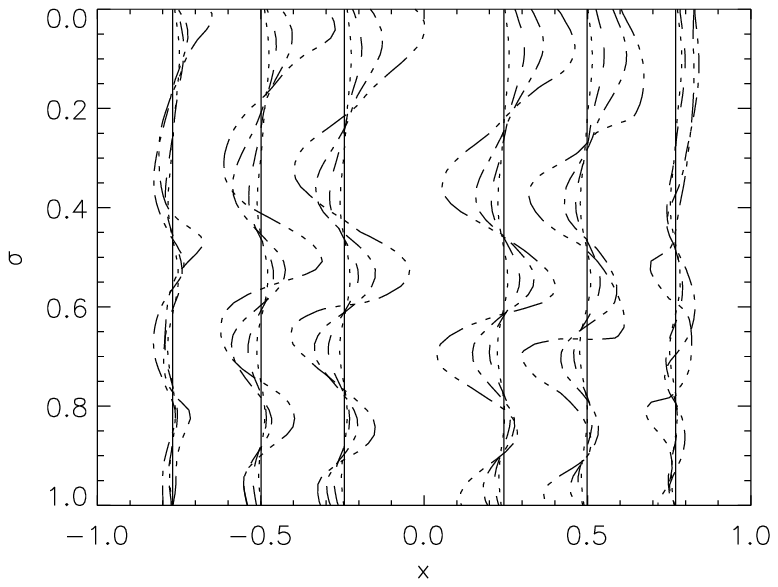}
\caption{Simulation of the Tayler instability (section
  \ref{subsec:tayler}). The lines show the positions of co-moving
  fluid surfaces as they intersect the $y=0$ cm plane at five
  different times: $\omega_{\rm A}t$=0, 1.55, 2.10, 2.63 and 3.20,
  represented respectively by the solid, dotted, dashed, dot-dashed
  and dot-dot-dot-dashed lines. The horizontal extent of the
  computational domain in both $x$ and $y$ is from $-2$ to $+2$ cm;
  only the central part of the $y=0$ cm plane is plotted here as
  nothing is happening towards the edges of the box.}
\label{fig:tayler}
\end{figure}

This is an instability of a toroidal magnetic field \citep{Tayler:1957,Tayler:1973}. The free energy source is the field itself and the energy is released by an interchange of fluid with weaker toroidal magnetic field $B_\phi$ with fluid containing stronger toroidal field at greater cylindrical radius $\varpi$. In a field given in the usual cylindrical coordinate notation by $B_\phi = B_0 \varpi/\varpi_0$ we expect the $m=1$ azimuthal mode to be unstable and the growth rate to be roughly equal to the Alfv\'en frequency given by $\omega_{\rm A}
\equiv v_{\rm A}/\varpi=B_0/(\varpi_0\sqrt{4\pi\rho})$. The instability can be modelled in a square computational box containing a magnetic field of the form $B_\phi = B_0 (\varpi/\varpi_0) / \{1+\exp[(\varpi-\varpi_0)/\Delta\varpi]\}$, the latter function simply being a smooth taper so that the field goes towards zero at the edge of the box. We use the same boundary conditions as in the previous case.

The horizontal size of the box is 4x4 cm$^2$ and we set $\varpi_0=4/3$
cm, $\Delta\varpi=0.12$ cm, $B_0=0.1$ G, the temperature at the
beginning is uniform and of value $RT=1$ erg g$^{-1}$ and we set $g=1$
cm s$^{-2}$, so that the scale height $H_p=1$ cm. The vertical extent
of the model is $0.01H_p$, which means that all vertical wavelengths
are expected to be unstable -- a strong stratification stabilizes the
longer wavelengths. A resolution 72x72x72 is used. The code
successfully reproduces the instability at all expected wavelengths,
and the growth rate measured corresponds to that expected (Fig.
\ref{fig:tayler}). Finally, the r.m.s. value of ${\bf
  \nabla^\ast}.{\bf B^\ast}$ (see equation \ref{equ:magfieldstar}) is
at most $2\times 10^{-5}$ of the r.m.s. of $B_z$ or of the r.m.s. of
$B_x dz/dx$, confirming again the good conservation of ${\bf
  \nabla^\ast}.{\bf B^\ast}$.

\section{Summary}\label{sec:conc}

We have described a numerical magnetohydrodynamic scheme designed to model
phenomena in gravitationally stratified fluids.  This
scheme uses the $\sigma$-coordinate system, a system which basically
employs pressure as the vertical coordinate. In order to do this the
code assumes hydrostatic equilibrium in the vertical direction.  Our code
is tailored for problems that fulfil the following conditions.  Firstly,
the fluid under consideration should have
strong gravitational stratification, with much greater length
scales in the horizontal than in the vertical direction (perhaps
greater than the scale height $H_{\rm p}$).  Secondly, the timescales
of interest should be longer than the vertical acoustic timescale
($H_{\rm p}/c_{\rm s}$). Finally, in the magnetic case, a high
plasma-$\beta$ is required so that Alfv\'en-wave propagation in the
vertical direction does not limit the timestep.

The code has been successfully validated.  It is capable of
reproducing very different phenomena like the Kelvin-Helmholtz and the
inverse entropy gradient instabilities, waves and shocks, the Rossby
adjustment problem as well as the propagation of Alfv\'en waves and
the Tayler instability. The code converges well to the analytic
solutions and conserves mass, energy and momentum very accurately.

This demonstrates our numerical scheme to be both highly flexible and
the natural choice for many astrophysical contexts such as planetary
atmospheres, stellar radiative zones, as well as neutron star
atmospheres. We have already used it to investigate flame propagation
in Type-I X-ray bursts on neutron stars: results from this study will
be presented elsewhere (Cavecchi et al.\ in prep.).

{\it Acknowledgements.} The authors would like to thank Evghenii Gaburov, Yuri Levin, Jonathan Mackey, Henk Spruit and Anna Watts for useful discussions and assistance. YC is funded by NOVA.


\bibliography{ms}

\label{lastpage}

\end{document}